\documentclass[12pt]{article}

\renewcommand{\thefootnote}{\fnsymbol{footnote}}

\usepackage{amsbsy,amssymb,latexsym,amsfonts,amsmath}
\usepackage{mathrsfs}
\usepackage{cite}
\usepackage{bm}
\usepackage{here}
\usepackage{comment}
\usepackage{amsmath}
\usepackage{cases}
\usepackage {empheq}
\usepackage{color}
\usepackage[pdftex]{graphicx}
\pdfoutput=1
\usepackage{braket}
\usepackage{slashed}
\usepackage{cancel}
\usepackage{enumerate}
\numberwithin{equation}{section}
\makeatletter
\def\EqNumText{\refstepcounter{equation}\cdots\tagform@\theequation}%
\makeatother
\usepackage{subcaption}

\newcommand{\bel}[1]{\begin{equation*}\label{#1}}                     
\newcommand{\bal}[1]{\begin{eqnarray}\label{#1}}                     
\newcommand{\be}{\begin{equation}}
\newcommand{\ee}{\end{equation}}

\begin{document}
%
%
\begin{titlepage}
\begin{flushright}
\normalsize
~~~~
February 13, 2023\\
\end{flushright}
	
\vspace{15pt}
	
\begin{center}
{\Large Interpolation and Exponentially Suppressed Cosmological Constant in Non-Supersymmetric Heterotic Strings with General $\mathbb{Z}_{2}$ Twists} \\
\end{center}
	
\vspace{20pt}
	
\begin{center}
{ Yuichi Koga\footnote{e-mail: d21sa001@st.osaka-cu.ac.jp}    }\\
%
\vspace{30pt}
%

\it Department of Physics, Graduate School of Science,\\
Osaka Metropolitan University\\
\vspace{5pt}
		
\vspace{5pt}
		
3-3-138, Sugimoto, Sumiyoshi-ku, Osaka, 558-8585, Japan \\

\end{center}
%
\vspace{15pt}
\begin{center}
Abstract\\
\end{center}
	
We study general non-supersymmetric heterotic string models, including so-called interpolating models, $d$-dimensionally compactified with the arbitrary number of freely acting $\mathbb{Z}_{2}$ twisted directions. Taking the limits of the compactified radii to zero and infinity (the endpoint limits), we show some examples of the various interpolation patterns in the $d=2$ (8-dimensional) case. In the region where supersymmetry is asymptotically restored, we derive the formula for the one-loop cosmological constant of $(10-d)$ dimensional non-supersymmetric heterotic string models with general $\mathbb{Z}_{2}$ twists, which does not depend on all the other endpoints and find out the points in the moduli space where the cosmological constant is exponentially suppressed. The moduli stability of the cosmological constant is also analyzed.
	
	
\vfill

\end{titlepage}

\renewcommand{\thefootnote}{\arabic{footnote}}
\setcounter{footnote}{0}
\tableofcontents

\section{Introduction}
In high-energy physics, it is one of the most critical problems to determine the scale of supersymmetry breaking. Recent experiments show that there is no evidence for supersymmetry at the accessible energy scale. It is worth considering the top-down scenario that supersymmetry is already broken at the Planck/string scale. 

Most of the non-supersymmetric string theories include a tachyon in their spectrum, but tachyon-free models without supersymmetry do also exist. In 10-dimensions, we have $SO(16)\times SO(16)$ heterotic model originally constructed in \cite{Dixon:1986iz, Alvarez-Gaume:1986ghj}, and tachyon-free string models without supersymmetry in four or general dimensions were constructed\cite{Ginsparg:1986wr, Kawai:1986ah}. In the top-down approach from non-supersymmetric string theories without tachyons, however, one faces a crucial problem: the vacuum energy density, i.e., cosmological constant, is extremely large. This raises the issue of vacuum instability as well as how to avoid conflict from observations. 

In this paper, we focus on the string models constructed by orbifolding with freely acting $\mathbb{Z}_{2}$ twists in which the cosmological constant can be exponentially suppressed\cite{Itoyama:1986ei, Itoyama:1987rc, Itoyama:2021itj, Itoyama:2021kxp, Itoyama:2019yst, Itoyama:2020ifw,Itoyama:2021fwc, Blum:1997gw, Abel:2015oxa, Aaronson:2016kjm, Abel:2017rch,Florakis:2016ani,Faraggi:2009xy,Faraggi:2021mws, Kounnas:2016gmz,Kounnas:2017mad, Coudarchet:2018ztz, Angelantonj:2019gvi, Abel:2020ldo} (other non-supersymmetric string models with small or vanishing cosmological constants have been proposed \cite{Moore:1987ue, Taylor:1987uv, Balog:1988dt, Dienes:1990ij, Dienes:1990qh, Kachru:1998hd, Kachru:1998pg, Kachru:1998yy, Satoh:2015nlc, Sugawara:2016lpa, Aoyama:2021kqa, Satoh:2021nfu}). This construction is a stringy version of the Scherk-Scwharz compactifications which breaks supersymmetry \cite{Scherk:1978ta, Rohm:1983aq, Kounnas:1989dk}. In the 9-dimensional models based on it, a 10-dimensional supersymmetric model can be obtained by taking the limits of compactified radii to zero or infinity in the specific choices of $\mathbb{Z}_{2}$ twists. Among these, so-called interpolating models have especially attracted much attention, which connect superstrings and non-supersymmetric ones in 10 dimensions and their interpolation properties are related to the target space duality of non-supersymmetric strings. It is shown that in 9-dimensional interpolating heterotic models, as the dimensionless radius $R$ of a circle goes to large (supersymmetry is asymptotically restored), the leading contribution of the one-loop cosmological constant can be evaluated as \cite{Itoyama:1986ei, Itoyama:1987rc}
\begin{align}\label{9d CC}
\Lambda^{(9)}=(n_{F}-n_{B})\xi R^{-9} + \mathcal{O}(e^{-R}),
\end{align}
where $n_{F}$ and $n_{B}$ represent the number of degrees of freedom of massless fermions and bosons, respectively, and $\xi$ is a computable positive constant. From (\ref{9d CC}), we find that the cosmological constant is exponentially suppressed in the 9-dimensional interpolating heterotic models when the degeneracy between boson and fermion is realized at the massless level. The $9$-dimensional interpolating heterotic models are investigated with (full) Wilson-line moduli \cite{Itoyama:2019yst, Itoyama:2020ifw}, and extended to $d$-dimensionally compactified ones in which only $X^{9}$-direction is twisted\cite{Itoyama:2021fwc}. We consider the further generalization to the $d$-dimensionally compactified models, including interpolating models, where the number of $\mathbb{Z}_{2}$ twisted directions is arbitrary. In bosonic constructions, the $\mathbb{Z}_{2}$ twist and the target space duality in non-supersymmetric strings are characterized by the shift vector $\delta$, half of which lives in the 16-dimensional even and self-dual lattice (Narain lattice)\cite{Dixon:1986iz, Ginsparg:1986wr, Itoyama:2021itj}. The components of $\delta$ determine the compactified directions where supersymmetry can be restored, as we will discuss later.

This paper is organized as follows. In section 2, we briefly review how to construct one-loop partition functions of non-supersymmetric heterotic models $d$-dimensionally compactified with general $\mathbb{Z}_{2}$ twists. In section 3, we investigate the behavior of the partition functions at the limits of the compactified radii to zero and infinity (which we call the endpoint limits) and show some examples of interpolations in $d=2$ (8-dimensional models). In section 4, the massless spectra with general $\mathbb{Z}_{2}$ twists are studied. In section 5, we calculate the one-loop cosmological constant in the region where supersymmetry is asymptotically restored. Under the assumption of the Wilson lines, we derive the formula to which we generalize (\ref{9d CC}) and find the configurations of Wison lines that give the exponential suppression. The Wilson-line moduli stability of the cosmological constant is also analyzed. This formula only depends on the choice of supersymmetric endpoint model, not on all the other ones. Finally, we conclude this paper in section 6. In appendix A, we summarize the $SO(2n)$ characters used to express the partition functions.
\section{Construction}\label{non-susy string}
In this section, we briefly review one way of constructing the non-supersymmetric heterotic string theories by $\mathbb{Z}_{2}$ twisted compactifications, based on \cite{Dixon:1986iz, Ginsparg:1986wr, Itoyama:2021itj}.
\subsection{Toroidal model}\label{Toroidal model}
Let us start from a toroidal model $d$-dimensionally compactified in which supersymmetry is maximally preserved. The one-loop partition function of the heterotic string models compactified on $T^d$ can be written as
\begin{align}\label{toroidal model}
Z^{T^d}=Z_{B}^{(8-d)}(\bar{V}_{8}-\bar{S}_{8})Z_{\Gamma^{16+d,d}}.
\end{align}
Here $Z_{B}^{(8-d)}=\tau_{2}^{-\frac{8-d}{2}}\left(\eta\bar{\eta} \right)^{-(8-d)}$ is the contribution from the bosonic part, $\bar{V}_{8}-\bar{S}_{8}$ is from the fermonic part and $Z_{\Gamma^{16+d,d}}$ is from the internal part;
\begin{align}
Z_{\Gamma^{16+d,d}}=\eta^{-(16+d)}\bar{\eta}^{-d}\sum_{p\in \Gamma^{16+d,d}}q^{\frac{1}{2}p_{L}^{2}}\bar{q}^{\frac{1}{2}p_{R}^{2}},
\end{align}
where $q=e^{2\pi i \tau}$, $\eta(\tau)$ is the Dedekind eta function and $\left(O_{8}, V_{8},S_{8},C_{8}\right)$ denotes a set of $SO(8)$ characters defined in appendix \ref{appendixA}. $\Gamma^{16+d,d}$ is a Narain lattice, an even self-dual lattice with Lorentzian signature $(16+d,d)$. Using the generalized vierbein $\mathcal{E}$ of the Narain lattice which is expressed as a $(16+2d)\times (16+2d)$ matrix, an element $p$ of $\Gamma^{16+d,d}$ is written as
\begin{align}
p=Z\mathcal{E},
\end{align}
where  $Z=\left(q,m,n \right)\in \mathbb{Z}^{16}\otimes \mathbb{Z}^d\otimes \mathbb{Z}^d$ is a $(16+2d)$-dimensional row vector with integer components. We define the inner product of $p_{1}=Z_{1}\mathcal{E}$ and $p_{2}=Z_{2}\mathcal{E}$ as
\begin{align}\label{inner product}
p_{1}\cdot p_{2}=Z_{1}\mathcal{E}\eta\mathcal{E}^{t}Z_{2}^{t}=Z_{1}JZ_{2}^{t},
\end{align}
where $\eta=diag\left(\boldsymbol{1}_{16+d},-\boldsymbol{1}_{d} \right)$ and $J=\mathcal{E}\eta\mathcal{E}^t$ is called the Narain metric. We can choose a Narain metric as
\begin{align}
J =\left(\begin{array}{ccc}
	g_{16} & 0 & 0\\
	0 & 0 & \boldsymbol{1}_{d}\\
	0 & \boldsymbol{1}_{d} & 0
\end{array}				\right),~~~g_{16}=\alpha_{16}\alpha_{16}^{t}
\end{align}
where $\alpha_{16}$ denotes a set of the basis of a 16-dimensional even self-dual Euclidean lattice $\Gamma^{16}$.

A vierbein $\mathcal{E}$ of the Narain lattice $\Gamma^{16+d,d}$ depends on a set of parameters, moduli. In the heterotic models $d$-dimensional toroidal compactified, there are $(16+d)\times d$ moduli,  that is, a metric $G$ of the compactification lattice, an anti-symmetric two-form $B$  and $16d$ Wilson lines $A$. In the special case of a rectangular $d$-torus, the internal circles are all perpendicular, and the internal metric $G$ is diagonal, resulting in
\begin{align}
G_{ij}=R^{2}_{i}\delta_{ij},
\end{align}
where $i=1,2,\ldots, d$ and $R_{i}$ is the radius of the $i$-th compactified circle normalized to be dimensionless by using the string lengthscale $\sqrt{\alpha'}$. Then the (dimensionless) internal momenta $p=(\ell_{L},p_{L},p_{R})$ is explicitly written as\cite{Narain:1986am}
\begin{subequations}\label{internal momenta}
\begin{align}
	\ell_{L}^{I} &= \pi^{I} - m^{i}A_{i}^{I}, \\
	p_{Li}&=  \frac{1}{\sqrt{2}R_{i}}\left( \pi\cdot A_{i} +n_{i}+ m^{j}\left( G_{ij}+B_{ij} -\frac{1}{2}A_{i}\cdot A_{j} \right)  \right) ,\\
	p_{Ri}&=  \frac{1}{\sqrt{2}R_{i}}\left( \pi\cdot A_{i} +n_{i}- m^{j}\left( G_{ij}-B_{ij}  +\frac{1}{2}A_{i}\cdot A_{j} \right)  \right),
\end{align}
\end{subequations}
where $I=1,\ldots,16$ and $\pi=q\alpha_{16}$ lives in $\Gamma^{16}$. Here we also define $\pi\cdot A_{i}=\sum_{I}\pi^{I}A_{i}^{I}$ and $A_{i}\cdot A_{j}=\sum_{I}A_{i}^{I}A_{j}^{I}$. Note that $m$ and $n$ label the winding number and the Kaluza-Klein momentum, respectively.

\subsection{Non-supersymmetric model}
The non-supersymmetric model is constructed by orbifolding the toroidal model by a $\mathbb{Z}_2$ shift action expressed as $(-1)^{F}\alpha$. Here $F$ is the spacetime fermion number, and $\alpha$ gives an eigenvalue $e^{2\pi i\delta\cdot p}$ for a state with an internal momentum $p$, where $\delta$ is a shift-vector in the Narain lattice such that $2\delta\in \Gamma^{16+d,d}$. For convenience, we split $\Gamma^{16+d,d}$ into two subsets $\Gamma^{16+d,d}_{+}$ and $\Gamma^{16+d,d}_{-}$ as follows:
\begin{align}\label{Gammapm}
	\Gamma_{+}^{16+d,d}(\delta)=\left\lbrace \left.  p\in \Gamma^{16+d,d}~\right|  \delta\cdot p\in \mathbb{Z} \right\rbrace,~~
	\Gamma_{-}^{16+d,d}(\delta)=\left\lbrace \left.  p\in \Gamma^{16+d,d} ~\right| \delta\cdot p \in \mathbb{Z} +\frac{1}{2}\right\rbrace.
\end{align}
Since $2\delta\in \Gamma^{16+d,d}$, the shift vector $\delta$ is written as
\begin{align}
	\delta=\frac{1}{2}\hat{Z}\mathcal{E},
\end{align}
for a certain integer vector $\hat{Z}=\left(\hat{q},\hat{m},\hat{n} \right)\in \mathbb{Z}^{16}\otimes \mathbb{Z}^d\otimes \mathbb{Z}^d$. Note that if two choices $\hat{Z}_{1}$ and $\hat{Z}_{2}$ satisfy $\hat{Z}_{1}=\hat{Z}_{2}~(\text{mod 2})$, then these give the same splitting of the Narain lattice by definition \eqref{Gammapm} of $\Gamma_{\pm}^{16+d,d}$. This means that we can consider the choices of $\hat{Z}$ where each of the components takes either 0 or 1,  excluding $\hat{Z}=\left( 0^{16+2d}\right)$. We also note that from the definition (\ref{inner product}), the inner product $\delta\cdot p$ can be written as
\begin{align}\label{pdelta}
	\delta\cdot p=\frac{1}{2}\hat{Z}J Z^{t}=\frac{1}{2}\left( \hat{\pi}\cdot\pi+\hat{m}n^{t}+\hat{n}m^{t} \right),
\end{align}
where we define $\hat{\pi}=\hat{q}\alpha_{16}$. The non-supersymmetric heterotic models are characterized by $\hat{\pi}, \hat{m},\hat{n}$.

The action $\alpha$ to a state with an internal momentum $p$ is expressed as
\begin{equation}
	\alpha\ket{p}=
	\begin{cases}
		+\ket{p} & \text{for}~~p\in\Gamma_{+}^{16+d,d}, \\
		-\ket{p} & \text{for}~~p\in\Gamma_{-}^{16+d,d}.
	\end{cases}
\end{equation}
The partition function (\ref{toroidal model}) projected by $\left(1+(-1)^{F}\alpha\right)/2$ is written as
\begin{align}
	Z_{B}^{(8-d)}\left(\bar{V}_{8}Z_{\Gamma_{+}^{16+d,d}}-\bar{S}_{8}Z_{\Gamma_{-}^{16+d,d}}\right).
\end{align}
We must impose modular invariance on the one-loop partition function\footnote{The modular transformations of the $SO(2n)$ characters are summarized in appendix \ref{appendixA}.}. It requires $\delta^{2}$ to be an integer and twisted sectors to be added. From the former condition, $\hat{Z}$ satisfies $\hat{Z}J\hat{Z}^t=0~(\text{mod 4})$. In other words, the constraint for $\delta^{2}$ is written as
\begin{align}\label{delta condition}
	\left| \hat{\pi}\right| ^2+2\hat{m}\hat{n}^{t}=0~~(\text{mod 4}),
\end{align}
where we define $|\hat{\pi}|^{2}=\hat{\pi}\hat{\pi}^{t}$. Orbifolding the toroidal model (\ref{toroidal model}) by $(-1)^{F}\alpha$ that depends on $\hat{Z}$ satisfying (\ref{delta condition}), the partition function of non-supersymmetric models is given as
\begin{align}\label{non-susy hetero}
	Z^{\cancel{SUSY}}_{(\hat{Z})}=Z_{B}^{(8-d)}\left\lbrace \bar{V}_{8} Z_{\Gamma^{16+d,d}_{+}}-\bar{S}_{8} Z_{\Gamma^{16+d,d}_{-}}
	+\bar{O}_{8} Z_{\Gamma^{16+d,d}_{\pm}+\delta}-\bar{C}_{8}  Z_{ \Gamma^{16+d,d}_{\mp}+\delta} \right\rbrace,
\end{align}
where the upper sign of $Z_{\Gamma^{d_{L},d_{R}}_{\pm}+\delta}$ applies for $\delta^2$ odd and the lower for $\delta^2$ even in the twisted sectors. The internal part $Z_{\Gamma^{d_{L},d_{R}}_{\pm}+\delta}$ is expressed as
\begin{align}
	Z_{\Gamma_{\pm}^{16+d,d}+\delta}
	=\eta^{-(16+d)}\bar{\eta}^{-d}\sum_{p\in \Gamma^{16+d,d}}\frac{1\pm e^{2\pi i \delta\cdot p}}{2}q^{\frac{1}{2}\left( p_{L}+\delta_{L} \right)^{2}}\bar{q}^{\frac{1}{2}\left( p_{R}+\delta_{R}\right)^{2}}.
\end{align}
Setting $d=0$, we can obtain the 10-dimensional non-supersymmetric heterotic models with starting points being the heterotic $Spin(32)/\mathbb{Z}_{2}$ and $E_{8}\times E_{8}$ models, as shown in Table \ref{table1}\cite{Dixon:1986iz}. In this case, the shift vector $\delta=\hat{\pi}/2$ is one-half of an element of the $Spin(32)/\mathbb{Z}_{2}$ or $E_{8}\times E_{8}$ root lattice.
\begin{table}[h]
	\centering
	\begin{tabular}{|c||c|c|c|}  \hline
		Lattice & $\delta=\frac{\hat{\pi}}{2}$ & Gauge symmetry  \\ \hline \hline
		$Spin(32)/\mathbb{Z}_{2}$ & $\left(1,0^{15} \right)$ & $SO(32)$  \\ \hline
		$Spin(32)/\mathbb{Z}_{2}$ & $\left(\left( \frac{1}{2}\right) ^4,0^{12}\right)$ & $SO(24)\times SO(8)$ \\ \hline
		$Spin(32)/\mathbb{Z}_{2}$ & $\left( \left( \frac{1}{4}\right)^{16}\right)$ & $SU(16)\times U(1)$  \\ \hline
		$Spin(32)/\mathbb{Z}_{2}$ & $\left(\left( \frac{1}{2}\right) ^8,0^8\right)$ & $SO(16)\times SO(16)$  \\ \hline \hline
		$E_{8}\times E_{8}$ & $\left(1,0^7;0^8 \right)$ & $SO(16)\times E_8$   \\ \hline
		$E_{8}\times E_{8}$ & $\left(\left( \frac{1}{2}\right) ^2,0^6;\left( \frac{1}{2}\right) ^2,0^6\right)$ & $\left( E_{7}\times SU(2)\right)^2 $   \\ \hline
		$E_{8}\times E_{8}$ & $\left(1,0^7;1,0^7\right) $ & $SO(16)\times SO(16)$   \\ \hline
	\end{tabular}
	\caption{The 10-dimensional non-supersymmetric heterotic models constructed from the $Spin(32)/\mathbb{Z}_{2}$ and $E_{8}\times E_{8}$ lattice by the shift vector $\delta$ and the realized gauge symmetries.}\label{table1}
\end{table}
\section{Endpoint limit and interpolation}\label{interpolation}
We study the behavior of the non-supersymmetric heterotic models (\ref{non-susy hetero}) which satisfy the condition (\ref{delta condition}) in the endpoint limits ($R_{i}\to \infty$ and $R_{i}\to 0$) with the moduli $A=B=0$. In this section, we only focus on the $d=2$ case, but the analysis can be easily generalized to the $d>2$ case. From the fact that the states with $m^{i}=0$ ($n_{i}=0$) only contribute as $R_{i}\to\infty$ ($R_{i}\to 0$) for each $i$-th direction, we show the various patterns of interpolation as examples in $d=2$. We find that the interpolation in $d=2$ can be recognized as combinations of those in $d=1$. Note that in $d$-dimensional compactified case, there are $2^{2d}$ classes since $\hat{m}^{i}$ and $\hat{n}_{i}$ are taken to be $0$ or $1$.

\subsection{Review of $d=1$ case}\label{d=1 case}
As preparation for the $d=2$ case, we review the endpoint limits in the $d=1$ case. In the 9-dimensional models, there are $2^{2}=4$ classes by the possible choices of $\hat{Z}$, as studied in \cite{Itoyama:2021itj}.
\begin{enumerate}
	\item $\left|\hat{\pi} \right|^2=0~(\text{mod 4}),~~\left(\hat{m};\hat{n} \right) =(0;0)$:\\
	In this class, the inner product (\ref{pdelta}) is written as 
	\begin{align}
		\delta\cdot p=\frac{1}{2} \hat{\pi}\cdot\pi.
	\end{align}
	From the definition (\ref{Gammapm}), $\Gamma_{\pm}^{17,1}$ and $\Gamma_{\pm}^{17,1}+\delta$ in this class are written as the following sets:
	\begin{align}
		&\Gamma_{\pm}^{17,1}=\left\lbrace p=Z\tilde{\mathcal{E}}\left|  \left(\pi,m,n \right)\in \left(\Gamma^{16}_{\pm},\mathbb{Z},\mathbb{Z} \right)  \right.  \right\rbrace,\\
		&\Gamma_{\pm}^{17,1}+\delta=\left\lbrace p=Z\tilde{\mathcal{E}}\left|\left(\pi,m,n \right)\in \left(\Gamma^{16}_{\pm}+\frac{\hat{\pi}}{2},\mathbb{Z},\mathbb{Z} \right) \right.  \right\rbrace,
	\end{align}
	where $\Gamma_{+}^{16}(\hat{\pi})$ and $\Gamma_{-}^{16}(\hat{\pi})$ are defined as
	\begin{align}
		\Gamma_{+}^{16}(\hat{\pi})=\left\lbrace \left.  \pi\in \Gamma^{16}~\right|  \hat{\pi}\cdot \pi\in 2\mathbb{Z} \right\rbrace,~~~~~
		\Gamma_{-}^{16}(\hat{\pi})=\left\lbrace \left.  \pi\in \Gamma^{16}~\right|  \hat{\pi}\cdot \pi\in 2\mathbb{Z}+1 \right\rbrace.
	\end{align}
	Since the states with $m^{1}=0$ ($n_{1}=0$) only contribute as $R_{1}\to\infty$ ($R_{1}\to 0$), then we find the behavior of $Z_{\Gamma_{\pm}^{17,1}}$ and $Z_{\Gamma_{\pm}^{17,1}+\delta}$ with Wilson-line moduli $A=0$ in the endpoint limits:
	\begin{align}
		&Z_{\Gamma_{\pm}^{17,1}}\xrightarrow{R_1 \to \infty} \frac{R_{1}}{\sqrt{\tau_{2}}}\left(\eta\bar{\eta} \right)^{-1} Z_{\Gamma_{\pm}^{16}},~~~Z_{\Gamma_{\pm}^{17,1}+\delta}\xrightarrow{R_1 \to \infty} \frac{R_{1}}{\sqrt{\tau_{2}}}\left(\eta\bar{\eta} \right)^{-1}Z_{\Gamma_{\pm}^{16}+\frac{\hat{\pi}}{2}},\\
		&Z_{\Gamma_{\pm}^{17,1}}\xrightarrow{R_1 \to 0} \frac{1}{R_{1}\sqrt{\tau_{2}}}\left(\eta\bar{\eta} \right)^{-1}Z_{\Gamma_{\pm}^{16}},~~~Z_{\Gamma_{\pm}^{17,1}+\delta}\xrightarrow{R_1 \to 0} \frac{1}{R_{1}\sqrt{\tau_{2}}}\left(\eta\bar{\eta} \right)^{-1}Z_{\Gamma_{\pm}^{16}+\frac{\hat{\pi}}{2}},
	\end{align}
	where $Z_{\Gamma_{\pm}^{16}}$ and $Z_{\Gamma_{\pm}^{16}+\frac{\hat{\pi}}{2}}$ are defined as
	\begin{align}
		Z_{\Gamma_{\pm}^{16}}=\eta^{-16}\sum_{\pi\in\Gamma^{16}_{\pm}}q^{\frac{1}{2}|\pi|^2},~~~~Z_{\Gamma_{\pm}^{16}+\frac{\hat{\pi}}{2}}=\eta^{-16}\sum_{\pi\in\Gamma^{16}_{\pm}}q^{\frac{1}{2}\left| \pi+\frac{\hat{\pi}}{2}\right| ^2}.
	\end{align}
	In this class, both endpoint limits give the same 10-dimensional non-supersymmetric model constructed by the shift-vector $\delta=\hat{\pi}/2$. In other words, the 9-dimensional non-supersymmetric models in this class are obtained by compactifying the 10-dimensional non-supersymmetric ones shown in Table 1 on a circle.
	
	\item $\left|\hat{\pi} \right|^2=0~(\text{mod 4}),~~\left(\hat{m};\hat{n} \right) =(1;0)$:\\
	In this class, the inner product (\ref{pdelta}) is given by
	\begin{align}
		\delta\cdot p=\frac{1}{2}\left( \hat{\pi}\cdot\pi+n_{1} \right).
	\end{align}
	We can write $\Gamma_{\pm}^{17,1}$ and $\Gamma_{\pm}^{17,1}+\delta$ as 
	\begin{align}
		&\Gamma_{\pm}^{17,1}=\left\lbrace p=Z\tilde{\mathcal{E}}\left| \left(\pi,m,n \right)\in\left( \Gamma^{16}_{\pm},\mathbb{Z},2\mathbb{Z}\right) \right.  \right\rbrace\oplus
		\left\lbrace p=Z\tilde{\mathcal{E}}\left| \left(\pi,m,n \right)\in\left( \Gamma^{16}_{\mp},\mathbb{Z},2\mathbb{Z}+1\right)  \right.  \right\rbrace,\\
		&\Gamma_{\pm}^{17,1}+\delta=\left\lbrace p=Z\tilde{\mathcal{E}}\left| \left(\pi,m,n \right)\in\left( \Gamma^{16}_{\pm}+\frac{\hat{\pi}}{2},\mathbb{Z}+\frac{1}{2},2\mathbb{Z}\right)  \right.  \right\rbrace
		\nonumber\\&~~~~~~~~~~~~~~~
		\oplus\left\lbrace p=Z\tilde{\mathcal{E}} \left|  \left(\pi,m,n \right)\in\left( \Gamma^{16}_{\mp}+\frac{\hat{\pi}}{2},\mathbb{Z}+\frac{1}{2},2\mathbb{Z}+1\right) \right.  \right\rbrace.
	\end{align}
	The behavior of $Z_{\Gamma_{\pm}^{17,1}}$ and $Z_{\Gamma_{\pm}^{17,1}+\delta}$ in the endpoint limits are given by 
	\begin{align}
		&Z_{\Gamma_{\pm}^{17,1}}\xrightarrow{R_1 \to \infty} \frac{R_{1}}{\sqrt{\tau_{2}}}\left(\eta\bar{\eta} \right)^{-1}Z_{\Gamma^{16}},~~~Z_{\Gamma_{\pm}^{17,1}+\delta}\xrightarrow{R_1 \to \infty} 0,\\
		&Z_{\Gamma_{\pm}^{17,1}}\xrightarrow{R_1 \to 0} \frac{1}{R_{1}\sqrt{\tau_{2}}}\left(\eta\bar{\eta} \right)^{-1}Z_{\Gamma_{\pm}^{16}},~~~Z_{\Gamma_{\pm}^{17,1}+\delta}\xrightarrow{R_1 \to 0} \frac{1}{R_{1}\sqrt{\tau_{2}}}\left(\eta\bar{\eta} \right)^{-1}Z_{\Gamma_{\pm}^{16}+\frac{\hat{\pi}}{2}}.
	\end{align}
	The limit $R_{1}\to \infty$ makes supersymmetry asymptotically restored while the limit $R_{1}\to0$ gives the 10-dimensional non-supersymmetric models in Table 1. 
	
	\item $\left|\hat{\pi} \right|^2=0~(\text{mod 4}),~~\left(\hat{m};\hat{n} \right) =(0;1)$:\\
	In this class, we can write the inner product (\ref{pdelta}) as
	\begin{align}
		\delta\cdot p=\frac{1}{2}\left( \hat{\pi}\cdot\pi+m^{1} \right).
	\end{align}
	Then $\Gamma_{\pm}^{17,1}$ and $\Gamma_{\pm}^{17,1}+\delta$ are written as 
	\begin{align}
		&\Gamma_{\pm}^{17,1}=\left\lbrace p=Z\tilde{\mathcal{E}}\left| \left(\pi,m,n \right)\in\left( \Gamma^{16}_{\pm},2\mathbb{Z},\mathbb{Z}\right) \right.  \right\rbrace\oplus
		\left\lbrace p=Z\tilde{\mathcal{E}}\left| \left(\pi,m,n \right)\in\left( \Gamma^{16}_{\mp},2\mathbb{Z}+1,\mathbb{Z}\right)  \right.  \right\rbrace,\\
		&\Gamma_{\pm}^{17,1}+\delta=\left\lbrace p=Z\tilde{\mathcal{E}}\left| \left(\pi,m,n \right)\in\left( \Gamma^{16}_{\pm}+\frac{\hat{\pi}}{2},2\mathbb{Z},\mathbb{Z}+\frac{1}{2}\right)  \right.  \right\rbrace
		\nonumber\\&~~~~~~~~~~~~~~~
		\oplus\left\lbrace p=Z\tilde{\mathcal{E}} \left|  \left(\pi,m,n \right)\in\left( \Gamma^{16}_{\mp}+\frac{\hat{\pi}}{2},2\mathbb{Z}+1,\mathbb{Z}+\frac{1}{2}\right) \right.  \right\rbrace.
	\end{align}
	The behavior of $Z_{\Gamma_{\pm}^{17,1}}$ and $Z_{\Gamma_{\pm}^{17,1}+\delta}$ in the endpoint limits are 
	\begin{align}
		&Z_{\Gamma_{\pm}^{17,1}}\xrightarrow{R_{1} \to \infty} \frac{R_{1}}{\sqrt{\tau_{2}}}\left(\eta\bar{\eta} \right)^{-1}Z_{\Gamma^{16}_{\pm}},
		~~~Z_{\Gamma_{\pm}^{17,1}+\delta}\xrightarrow{R_{1} \to \infty} \frac{R_{1}}{\sqrt{\tau_{2}}}\left(\eta\bar{\eta} \right)^{-1}Z_{\Gamma^{16}_{\pm}+\frac{\hat{\pi}}{2}},\\
		&Z_{\Gamma_{\pm}^{17,1}}\xrightarrow{R_{1} \to 0} \frac{1}{R_{1}\sqrt{\tau_{2}}}\left(\eta\bar{\eta} \right)^{-1}Z_{\Gamma^{16}},~~~Z_{\Gamma_{\pm}^{17,1}+\delta}\xrightarrow{R_{1} \to 0} 0.
	\end{align} 
	In this class, the 10-dimensional non-supersymmetric heterotic models are produced in the limit $R_{1}\to\infty$, and the 10-dimensional supersymmetric heterotic ones are obtained in the limit $R_{1}\to 0$. 
	
	\item $\left|\hat{\pi} \right|^2=2~(\text{mod 4}),~~\left(\hat{m};\hat{n} \right) =(1;1)$:\\
	In this class, the inner product (\ref{pdelta}) is written as
	\begin{align}
		\delta\cdot p=\frac{1}{2}\left( \hat{\pi}\cdot\pi+m^{1}+n_{1} \right).
	\end{align}
	We can get $\Gamma_{\pm}^{17,1}$ and $\Gamma_{\pm}^{17,1}+\delta$ as 
	\begin{align}
		&\Gamma_{\pm}^{17,1}=
		\left\lbrace p=Z\tilde{\mathcal{E}}\left| \left(\pi,m,n \right)\in\left( \Gamma^{16}_{\pm},2\mathbb{Z},2\mathbb{Z}\right)    \right.  \right\rbrace\nonumber\\&~~~~~~~~~~
		\oplus\left\lbrace p=Z\tilde{\mathcal{E}}\left| \left(\pi,m,n \right)\in\left( \Gamma^{16}_{\pm},2\mathbb{Z}+1,2\mathbb{Z}+1\right)   \right.  \right\rbrace\nonumber\\&~~~~~~~~~~
		\oplus\left\lbrace p=Z\tilde{\mathcal{E}}\left| \left(\pi,m,n \right)\in\left( \Gamma^{16}_{\mp},2\mathbb{Z},2\mathbb{Z}+1\right)  \right.  \right\rbrace\nonumber\\&~~~~~~~~~~
		\oplus\left\lbrace p=Z\tilde{\mathcal{E}}\left| \left(\pi,m,n \right)\in\left( \Gamma^{16}_{\mp},2\mathbb{Z}+1,2\mathbb{Z}\right)  \right.  \right\rbrace,\\
		&\Gamma_{\pm}^{17,1}+\delta=
		\left\lbrace p=Z\tilde{\mathcal{E}}\left| \left(\pi,m,n \right)\in\left( \Gamma^{16}_{\pm}+\frac{\hat{\pi}}{2},2\mathbb{Z}+\frac{1}{2},2\mathbb{Z}+\frac{1}{2}\right)\right.  \right\rbrace
		\nonumber\\&~~~~~~~~~~~~~~
		\oplus\left\lbrace p=Z\tilde{\mathcal{E}}\left| \left(\pi,m,n \right)\in\left( \Gamma^{16}_{\pm}+\frac{\hat{\pi}}{2},2\mathbb{Z}-\frac{1}{2},2\mathbb{Z}-\frac{1}{2}\right) \right.  \right\rbrace
		\nonumber\\&~~~~~~~~~~~~~~
		\oplus\left\lbrace p=Z\tilde{\mathcal{E}}\left| \left(\pi,m,n \right)\in\left( \Gamma^{16}_{\mp}+\frac{\hat{\pi}}{2},2\mathbb{Z}+\frac{1}{2},2\mathbb{Z}-\frac{1}{2}\right) \right.  \right\rbrace
		\nonumber\\&~~~~~~~~~~~~~~
		\oplus\left\lbrace p=Z\tilde{\mathcal{E}}\left| \left(\pi,m,n \right)\in\left( \Gamma^{16}_{\mp}+\frac{\hat{\pi}}{2},2\mathbb{Z}-\frac{1}{2},2\mathbb{Z}+\frac{1}{2}\right) \right.  \right\rbrace.
	\end{align}	
	In the endpoint limits, $Z_{\Gamma_{\pm}^{17,1}}$ and $Z_{\Gamma_{\pm}^{17,1}+\delta}$ behave as follows:
	\begin{align}
		&Z_{\Gamma_{\pm}^{17,1}}\xrightarrow{R_{1} \to \infty} \frac{R_{1}}{\sqrt{\tau_{2}}}\left(\eta\bar{\eta} \right)^{-1}Z_{\Gamma^{16}},
		~~~Z_{\Gamma_{\pm}^{17,1}+\delta}\xrightarrow{R_{1} \to \infty} 0,\\
		&Z_{\Gamma_{\pm}^{17,1}}\xrightarrow{R_{1} \to 0} \frac{1}{R_{1}\sqrt{\tau_{2}}}\left(\eta\bar{\eta} \right)^{-1}Z_{\Gamma^{16}},~~~Z_{\Gamma_{\pm}^{17,1}+\delta}\xrightarrow{R_{1} \to 0} 0.
	\end{align} 
	In this class, supersymmetry is asymptotically restored in both endpoint limits though broken at finite values of $R_{1}$.
\end{enumerate}

We define $\Gamma_{\pm}^{17,1}|_{(k)}$ and $\Gamma_{\pm}^{17,1}+\delta|_{(k)}$ as the $\Gamma_{\pm}^{17,1}$ and $\Gamma_{\pm}^{17,1}+\delta$ in the class $(k)$ defined above, and we also define $\Gamma^{17,1}$ as
\begin{align}
	\Gamma^{17,1}=\left\lbrace p=Z\tilde{\mathcal{E}}\left|  \left(\pi,m,n \right)\in \left(\Gamma^{16},\mathbb{Z},\mathbb{Z} \right)  \right.  \right\rbrace.
\end{align}
The models in class (2) and class (3) are called interpolating models since they interpolate between two different higher-dimensional string models, originally constructed in \cite{Itoyama:1986ei, Itoyama:1987rc}. We find that $\hat{m}, \hat{n}=1$ implies that there is a  $\mathbb{Z}_{2}$ twist on a compactified circle: the limit of $R_{1}\to \infty ~(R_{1}\to 0)$ makes supersymmetry restored if $\hat{m}=1 ~(\hat{n}=1)$.

\subsection{Some examples of $d=2$ case}\label{d=2 case}
\begin{figure}[h]
	\begin{minipage}[b]{0.5\linewidth}
		\centering
		\includegraphics[keepaspectratio, scale=0.5]{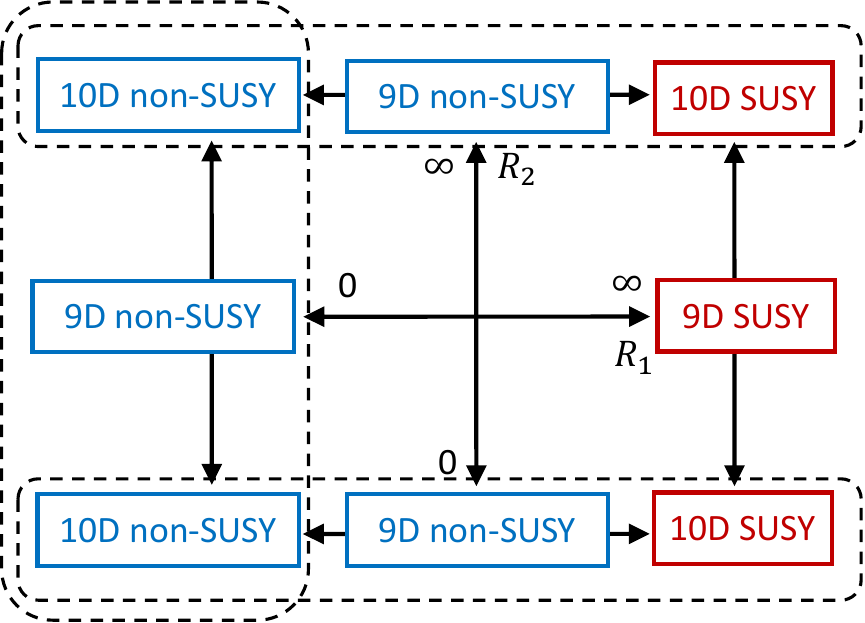}
		\subcaption{$\left|\hat{\pi} \right|^2=0~(\text{mod 4})$, $ \left(\hat{m};\hat{n} \right) =(1,0;0,0)$}\label{(1,0;0,0)model}
	\end{minipage}
	\vspace{2mm}
	\begin{minipage}[b]{0.5\linewidth}
		\centering
		\includegraphics[keepaspectratio, scale=0.5]{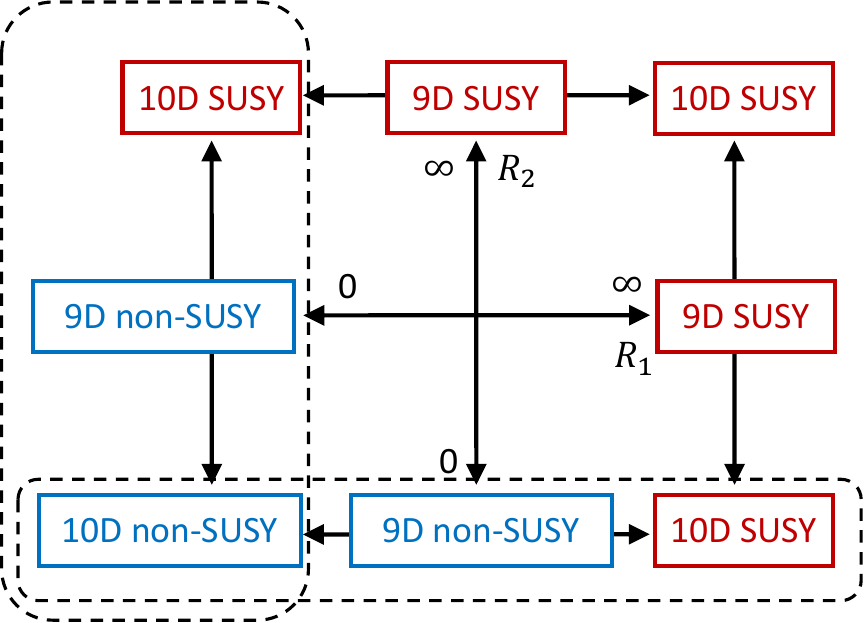}
		\subcaption{$\left|\hat{\pi} \right|^2=0~(\text{mod 4})$, $ \left(\hat{m};\hat{n} \right) =(1,1;0,0)$}\label{(1,1;0,0)model}
	\end{minipage}
	\vspace{2mm}
	\begin{minipage}[b]{0.5\linewidth}
		\centering
		\includegraphics[keepaspectratio, scale=0.5]{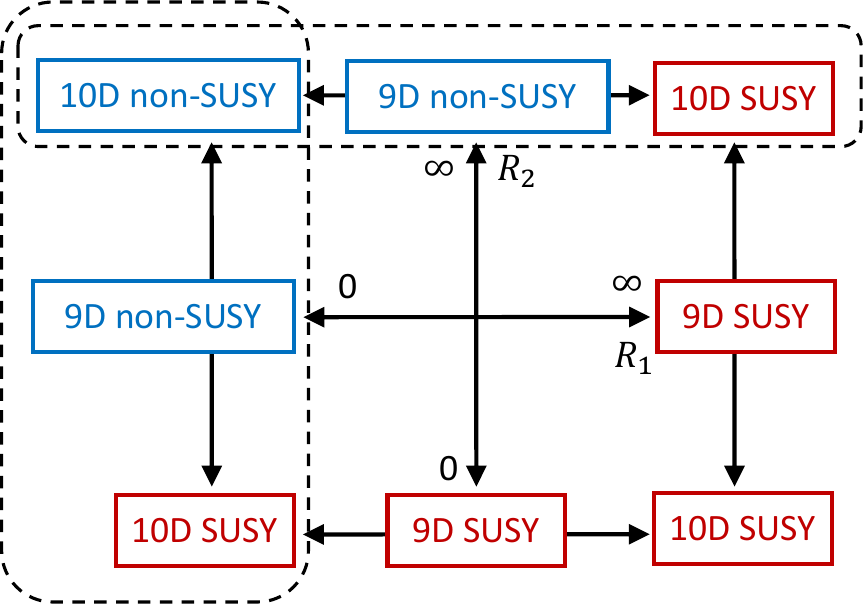}
		\subcaption{$\left|\hat{\pi} \right|^2=0~(\text{mod 4})$, $ \left(\hat{m};\hat{n} \right) =(1,0;0,1)$}\label{(1,0;0,1)model}
	\end{minipage}
	\vspace{2mm}
	\begin{minipage}[b]{0.5\linewidth}
		\centering
		\includegraphics[keepaspectratio, scale=0.5]{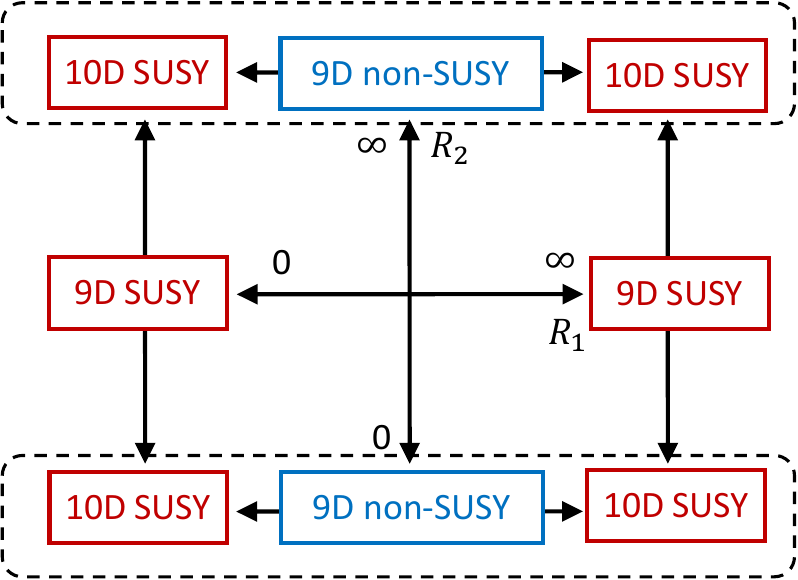}
		\subcaption{$\left|\hat{\pi} \right|^2=2~(\text{mod 4})$, $ \left(\hat{m};\hat{n} \right) =(1,0;1,0)$}\label{(1,0;1,0)model}
	\end{minipage}
	\vspace{-10mm}
	\caption{The illustrations on endpoint limits of the 8-dimensional non-supersymmetric heterotic models. The dotted lines in each diagram refer to the 9-dimensional non-supersymmetric heterotic models classified into four classes.}\label{d=2model}
\end{figure}

The 8-dimensional non-supersymmetric heterotic models are classified into $2^{4}=16$ classes. With $d=2$, more patterns of interpolation (endpoint limits) are shown than those in the $d=1$ case, in the limits of compactified radii to zero or infinity. Here we give the four concrete examples and illustrate the endpoint limits of these non-supersymmetric heterotic models in Fig. \ref{d=2model}. In each class, the interpolation patterns of $d=2$ case are identified as the combinations of those of $d=1$.
\begin{enumerate}[(a)]
	\item $\left|\hat{\pi} \right|^2=0~~(\text{mod 4}), ~~\left(\hat{m};\hat{n} \right) =(1,0;0,0)$:\\
	In this class, the inner product (\ref{pdelta}) is written as
	\begin{align}
		\delta\cdot p=\frac{1}{2}\left( \hat{\pi}\cdot\pi+n_{1} \right).
	\end{align}
	From this, $\Gamma_{\pm}^{18,2}$ and $\Gamma_{\pm}^{18,2}+\delta$ are written as 
	\begin{align}
		&\Gamma_{\pm}^{18,2}=\left\lbrace p=Z\tilde{\mathcal{E}}\left| \left(\pi,m,n \right)\in\left( \Gamma^{16}_{\pm},\mathbb{Z}^{2},2\mathbb{Z} \times \mathbb{Z}\right) \right.  \right\rbrace
		\nonumber\\&~~~~~~~~~~~
		\oplus \left\lbrace p=Z\tilde{\mathcal{E}}\left| \left(\pi,m,n \right)\in\left( \Gamma^{16}_{\mp},\mathbb{Z}^{2},(2\mathbb{Z}+1)\times \mathbb{Z}\right)  \right.  \right\rbrace,\\
		&\Gamma_{\pm}^{18,2}+\delta=\left\lbrace p=Z\tilde{\mathcal{E}}\left| \left(\pi,m,n \right)\in\left( \Gamma^{16}_{\pm}+\frac{\hat{\pi}}{2},(\mathbb{Z}+\frac{1}{2})\times \mathbb{Z},2\mathbb{Z} \times \mathbb{Z}\right)  \right.  \right\rbrace
		\nonumber\\&~~~~~~~~~~~~~~~
		\oplus \left\lbrace p=Z\tilde{\mathcal{E}}\left| \left(\pi,m,n \right)\in\left( \Gamma^{16}_{\mp}+\frac{\hat{\pi}}{2},(\mathbb{Z}+\frac{1}{2})\times \mathbb{Z},(2\mathbb{Z}+1)\times \mathbb{Z}\right)  \right.  \right\rbrace.
	\end{align}
	Let us study the behavior of partition functions in the endpoint limits ($R_{i}\to \infty$ and $R_{i}\to 0$) with the moduli $A=B=0$. We find the behavior of $Z_{\Gamma_{\pm}^{18,2}}$ is
	\begin{align*}
		&Z_{\Gamma_{\pm}^{18,2}}\xrightarrow{R_1 \to \infty} \frac{R_1}{\sqrt{\tau_{2}}}\left(\eta\bar{\eta} \right)^{-1} Z_{\Gamma^{17,1}}\xrightarrow{R_2 \to \infty} \frac{R_1 R_2}{\tau_{2}}\left(\eta\bar{\eta} \right)^{-2} Z_{\Gamma^{16}},\\
		&Z_{\Gamma_{\pm}^{18,2}}\xrightarrow{R_2 \to \infty} \frac{R_2}{\sqrt{\tau_{2}}}\left(\eta\bar{\eta} \right)^{-1} Z_{\Gamma_{\pm}^{17,1}|_{(2)}}\xrightarrow{R_1 \to \infty} \frac{R_1 R_2}{\tau_{2}}\left(\eta\bar{\eta} \right)^{-2} Z_{\Gamma^{16}},\\
		&Z_{\Gamma_{\pm}^{18,2}}\xrightarrow{R_1 \to 0}\frac{1}{R_1\sqrt{\tau_{2}}}\left(\eta\bar{\eta} \right)^{-1}Z_{\Gamma_{\pm}^{17,1}|_{(1)}}\xrightarrow{R_2 \to 0} \frac{1}{R_1 R_2\tau_{2}}\left(\eta\bar{\eta} \right)^{-2} Z_{\Gamma_{\pm}^{16}},\\
		&Z_{\Gamma_{\pm}^{18,2}}\xrightarrow{R_2 \to 0}\frac{1}{R_2\sqrt{\tau_{2}}}\left(\eta\bar{\eta} \right)^{-1}Z_{\Gamma_{\pm}^{17,1}|_{(2)}}\xrightarrow{R_1 \to 0} \frac{1}{R_1 R_2\tau_{2}}\left(\eta\bar{\eta} \right)^{-2} Z_{\Gamma_{\pm}^{16}},\\
		&Z_{\Gamma_{\pm}^{18,2}}\xrightarrow{R_1 \to \infty,R_2 \to 0} \frac{R_1}{R_2 \tau_{2}}\left(\eta\bar{\eta} \right)^{-2} Z_{\Gamma^{16}},\\
		&Z_{\Gamma_{\pm}^{18,2}}\xrightarrow{R_1 \to 0,R_2 \to \infty} \frac{R_2}{R_1 \tau_{2}}\left(\eta\bar{\eta} \right)^{-2} Z_{\Gamma_{\pm}^{16}},
	\end{align*}
	and that of $Z_{\Gamma_{\pm}^{18,2}+\delta}$ is
	\begin{align*}
		&Z_{\Gamma_{\pm}^{18,2}+\delta}\xrightarrow{R_1 \to \infty}0,\\
		&Z_{\Gamma_{\pm}^{18,2}+\delta}\xrightarrow{R_2 \to \infty} \frac{R_2}{\sqrt{\tau_{2}}}\left(\eta\bar{\eta} \right)^{-1} Z_{\Gamma_{\pm}^{17,1}+\delta|_{(2)}}\xrightarrow{R_1 \to \infty}0,\\
		&Z_{\Gamma_{\pm}^{18,2}+\delta}\xrightarrow{R_1 \to 0}\frac{1}{R_1\sqrt{\tau_{2}}}\left(\eta\bar{\eta} \right)^{-1}Z_{\Gamma_{\pm}^{17,1}+\delta|_{(1)}}\xrightarrow{R_2 \to 0} \frac{1}{R_1 R_2\tau_{2}}\left(\eta\bar{\eta} \right)^{-2} Z_{\Gamma_{\pm}^{16}+\frac{\hat{\pi}}{2}},\\
		&Z_{\Gamma_{\pm}^{18,2}+\delta}\xrightarrow{R_2\to 0}\frac{1}{R_2\sqrt{\tau_{2}}}\left(\eta\bar{\eta} \right)^{-1}Z_{\Gamma_{\pm}^{17,1}+\delta|_{(2)}}\xrightarrow{R_1 \to 0} \frac{1}{R_1 R_2\tau_{2}}\left(\eta\bar{\eta} \right)^{-2} Z_{\Gamma_{\pm}^{16}+\frac{\hat{\pi}}{2}},\\
		&Z_{\Gamma_{\pm}^{18,2}+\delta}\xrightarrow{R_1 \to 0,R_2 \to \infty} \frac{R_2}{R_1 \tau_{2}}\left(\eta\bar{\eta} \right)^{-2} Z_{\Gamma_{\pm}^{16}+\frac{\hat{\pi}}{2}}.
	\end{align*}
	The limit $R_{1} \to 0$ gives the 9-dimensional non-supersymmetric model, which belongs to the class (1). In contrast, both $R_{2} \to \infty$ and $R_{2} \to 0$ limits give the 9-dimensional non-supersymmetric model, which belongs to the class (2), so-called interpolating model: $R_{1} \to \infty$ gives a 10D superstring while $R_{1} \to 0$ gives a 10-dimensional non-supersymmetric string.
	
	\item $\left|\hat{\pi} \right|^2=0~~(\text{mod 4}), ~~\left(\hat{m};\hat{n} \right) =(1,1;0,0)$:\\
	In this class, the inner product (\ref{pdelta}) is written as
	\begin{align}
		\delta\cdot p=\frac{1}{2}\left( \hat{\pi}\cdot\pi+n_{1}+n_{2} \right).
	\end{align}
	Then $\Gamma_{\pm}^{18,2}$ and $\Gamma_{\pm}^{18,2}+\delta$ are written as 
	\begin{align}
		&\Gamma_{\pm}^{18,2}=\left\lbrace p=Z\tilde{\mathcal{E}}\left| \left(\pi,m,n \right)\in\left( \Gamma^{16}_{\pm},\mathbb{Z}^{2},2\mathbb{Z} \times 2\mathbb{Z}\right) \right.  \right\rbrace
		\nonumber\\&~~~~~~~~~~~
		\oplus \left\lbrace p=Z\tilde{\mathcal{E}}\left| \left(\pi,m,n \right)\in\left( \Gamma^{16}_{\pm},\mathbb{Z}^{2},(2\mathbb{Z}+1)\times (2\mathbb{Z}+1)\right)  \right.  \right\rbrace,\nonumber\\&~~~~~~~~~~~
		\oplus \left\lbrace p=Z\tilde{\mathcal{E}}\left| \left(\pi,m,n \right)\in\left( \Gamma^{16}_{\mp},\mathbb{Z}^{2},2\mathbb{Z}\times (2\mathbb{Z}+1)\right)  \right.  \right\rbrace,\nonumber\\&~~~~~~~~~~~
		\oplus \left\lbrace p=Z\tilde{\mathcal{E}}\left| \left(\pi,m,n \right)\in\left( \Gamma^{16}_{\mp},\mathbb{Z}^{2},(2\mathbb{Z}+1)\times 2\mathbb{Z}\right)  \right.  \right\rbrace,\\
		&\Gamma_{\pm}^{18,2}+\delta=\left\lbrace p=Z\tilde{\mathcal{E}}\left| \left(\pi,m,n \right)\in\left( \Gamma^{16}_{\pm},(\mathbb{Z}+\frac{1}{2})^{2},2\mathbb{Z} \times 2\mathbb{Z}\right) \right.  \right\rbrace
		\nonumber\\&~~~~~~~~~~~~~~~
		\oplus \left\lbrace p=Z\tilde{\mathcal{E}}\left| \left(\pi,m,n \right)\in\left( \Gamma^{16}_{\pm},(\mathbb{Z}+\frac{1}{2})^{2},(2\mathbb{Z}+1)\times (2\mathbb{Z}+1)\right)  \right.  \right\rbrace,\nonumber\\&~~~~~~~~~~~~~~~
		\oplus \left\lbrace p=Z\tilde{\mathcal{E}}\left| \left(\pi,m,n \right)\in\left( \Gamma^{16}_{\mp},(\mathbb{Z}+\frac{1}{2})^{2},2\mathbb{Z}\times (2\mathbb{Z}+1)\right)  \right.  \right\rbrace,\nonumber\\&~~~~~~~~~~~~~~~
		\oplus \left\lbrace p=Z\tilde{\mathcal{E}}\left| \left(\pi,m,n \right)\in\left( \Gamma^{16}_{\mp},(\mathbb{Z}+\frac{1}{2})^{2},(2\mathbb{Z}+1)\times 2\mathbb{Z}\right)  \right.  \right\rbrace,
	\end{align}
	The behavior of $Z_{\Gamma_{\pm}^{18,2}}$ in the endpoint limits is given as
	\begin{align*}
		&Z_{\Gamma_{\pm}^{18,2}}\xrightarrow{R_1 \to \infty} \frac{R_1}{\sqrt{\tau_{2}}}\left(\eta\bar{\eta} \right)^{-1} Z_{\Gamma^{17,1}}\xrightarrow{R_2 \to \infty} \frac{R_1 R_2}{\tau_{2}}\left(\eta\bar{\eta} \right)^{-2} Z_{\Gamma^{16}},\\
		&Z_{\Gamma_{\pm}^{18,2}}\xrightarrow{R_2 \to \infty} \frac{R_2}{\sqrt{\tau_{2}}}\left(\eta\bar{\eta} \right)^{-1} Z_{\Gamma^{17,1}}\xrightarrow{R_1 \to \infty} \frac{R_1 R_2}{\tau_{2}}\left(\eta\bar{\eta} \right)^{-2} Z_{\Gamma^{16}},\\
		&Z_{\Gamma_{\pm}^{18,2}}\xrightarrow{R_1 \to 0}\frac{1}{R_1\sqrt{\tau_{2}}}\left(\eta\bar{\eta} \right)^{-1}Z_{\Gamma_{\pm}^{17,1}|_{(2)}}\xrightarrow{R_2 \to 0} \frac{1}{R_1 R_2\tau_{2}}\left(\eta\bar{\eta} \right)^{-2} Z_{\Gamma_{\pm}^{16}},\\
		&Z_{\Gamma_{\pm}^{18,2}}\xrightarrow{R_2 \to 0}\frac{1}{R_2\sqrt{\tau_{2}}}\left(\eta\bar{\eta} \right)^{-1}Z_{\Gamma_{\pm}^{17,1}|_{(2)}}\xrightarrow{R_1 \to 0} \frac{1}{R_1 R_2\tau_{2}}\left(\eta\bar{\eta} \right)^{-2} Z_{\Gamma^{16}},\\
		&Z_{\Gamma_{\pm}^{18,2}}\xrightarrow{R_1 \to \infty,R_2 \to 0} \frac{R_1}{R_2 \tau_{2}}\left(\eta\bar{\eta} \right)^{-2} Z_{\Gamma^{16}},\\
		&Z_{\Gamma_{\pm}^{18,2}}\xrightarrow{R_1 \to 0,R_2 \to \infty} \frac{R_2}{R_1 \tau_{2}}\left(\eta\bar{\eta} \right)^{-2} Z_{\Gamma^{16}},
	\end{align*}
	and that of $Z_{\Gamma_{\pm}^{18,2}+\delta}$ as
	\begin{align*}
		&Z_{\Gamma_{\pm}^{18,2}+\delta}\xrightarrow{R_1 \to \infty}0,\\
		&Z_{\Gamma_{\pm}^{18,2}+\delta}\xrightarrow{R_2 \to \infty} 0,\\
		&Z_{\Gamma_{\pm}^{18,2}+\delta}\xrightarrow{R_1 \to 0}\frac{1}{R_1\sqrt{\tau_{2}}}\left(\eta\bar{\eta} \right)^{-1}Z_{\Gamma_{\pm}^{17,1}+\delta|_{(2)}}\xrightarrow{R_2 \to 0} \frac{1}{R_1 R_2\tau_{2}}\left(\eta\bar{\eta} \right)^{-2} Z_{\Gamma_{\pm}^{16}+\frac{\hat{\pi}}{2}},\\
		&Z_{\Gamma_{\pm}^{18,2}+\delta}\xrightarrow{R_2\to 0} \frac{1}{R_2\sqrt{\tau_{2}}}\left(\eta\bar{\eta} \right)^{-1}Z_{\Gamma_{\pm}^{17,1}+\delta|_{(2)}}\xrightarrow{R_1 \to 0} \frac{1}{R_1 R_2\tau_{2}}\left(\eta\bar{\eta} \right)^{-2} Z_{\Gamma_{\pm}^{16}+\frac{\hat{\pi}}{2}},\\
		&Z_{\Gamma_{\pm}^{18,2}+\delta}\xrightarrow{R_1 \to 0,R_2 \to \infty} 0,\\
		&Z_{\Gamma_{\pm}^{18,2}+\delta}\xrightarrow{R_2 \to 0,R_1 \to \infty} 0.
	\end{align*}
	We find that the limit $R_{1}(R_{2}) \to 0$ gives the 9-dimensional interpolating model, which belongs to the class (2): $R_{2} (R_{1}) \to \infty$ gives a 10-dimensional supersymmetric model while $R_{2} (R_{1}) \to 0$ gives a 10-dimensional non-supersymmetric model. 
	
	\item $\left|\hat{\pi} \right|^2=0~~(\text{mod 4}), ~~\left(\hat{m};\hat{n} \right) =(1,0;0,1)$:\\
	In this class, the inner product (\ref{pdelta}) is written as
	\begin{align}
		\delta\cdot p=\frac{1}{2}\left( \hat{\pi}\cdot\pi+m^{2}+n_{1} \right).
	\end{align}
	Then $\Gamma_{\pm}^{18,2}$ and $\Gamma_{\pm}^{18,2}+\delta$ are written as 
	\begin{align}
		&\Gamma_{\pm}^{18,2}=\left\lbrace p=Z\tilde{\mathcal{E}}\left| \left(\pi,m,n \right)\in\left( \Gamma^{16}_{\pm},\mathbb{Z} \times 2\mathbb{Z} ,2\mathbb{Z} \times \mathbb{Z}\right) \right.  \right\rbrace
		\nonumber\\&~~~~~~~~~~~
		\oplus \left\lbrace p=Z\tilde{\mathcal{E}}\left| \left(\pi,m,n \right)\in\left( \Gamma^{16}_{\pm},\mathbb{Z} \times (2\mathbb{Z}+1),(2\mathbb{Z}+1)\times \mathbb{Z}\right)  \right.  \right\rbrace\nonumber\\&~~~~~~~~~~~
		\oplus \left\lbrace p=Z\tilde{\mathcal{E}}\left| \left(\pi,m,n \right)\in\left( \Gamma^{16}_{\mp},\mathbb{Z} \times 2\mathbb{Z},(2\mathbb{Z}+1)\times \mathbb{Z}\right)  \right.  \right\rbrace\nonumber\\&~~~~~~~~~~~
		\oplus \left\lbrace p=Z\tilde{\mathcal{E}}\left| \left(\pi,m,n \right)\in\left( \Gamma^{16}_{\mp},\mathbb{Z} \times (2\mathbb{Z}+1),2\mathbb{Z}\times \mathbb{Z}\right)  \right.  \right\rbrace,\\
		&\Gamma_{\pm}^{18,2}+\delta=\left\lbrace p=Z\tilde{\mathcal{E}}\left| \left(\pi,m,n \right)\in\left( \Gamma^{16}_{\pm}+\frac{\hat{\pi}}{2},(\mathbb{Z}+\frac{1}{2})\times 2\mathbb{Z},2\mathbb{Z} \times (\mathbb{Z}+\frac{1}{2})\right) \right.  \right\rbrace
		\nonumber\\&~~~~~~~~~~~
		\oplus \left\lbrace p=Z\tilde{\mathcal{E}}\left| \left(\pi,m,n \right)\in\left( \Gamma^{16}_{\pm}+\frac{\hat{\pi}}{2},(\mathbb{Z}+\frac{1}{2}) \times (2\mathbb{Z}+1),(2\mathbb{Z}+1)\times (\mathbb{Z}+\frac{1}{2})\right)  \right.  \right\rbrace\nonumber\\&~~~~~~~~~~~
		\oplus \left\lbrace p=Z\tilde{\mathcal{E}}\left| \left(\pi,m,n \right)\in\left( \Gamma^{16}_{\mp}+\frac{\hat{\pi}}{2},(\mathbb{Z}+\frac{1}{2}) \times 2\mathbb{Z},(2\mathbb{Z}+1)\times (\mathbb{Z}+\frac{1}{2})\right)  \right.  \right\rbrace\nonumber\\&~~~~~~~~~~~
		\oplus \left\lbrace p=Z\tilde{\mathcal{E}}\left| \left(\pi,m,n \right)\in\left( \Gamma^{16}_{\mp}+\frac{\hat{\pi}}{2},(\mathbb{Z}+\frac{1}{2}) \times (2\mathbb{Z}+1),2\mathbb{Z}\times (\mathbb{Z}+\frac{1}{2})\right)  \right.  \right\rbrace.
	\end{align}
	In the endpoint limits, $Z_{\Gamma_{\pm}^{18,2}}$ behave as
	\begin{align*}
		&Z_{\Gamma_{\pm}^{18,2}}\xrightarrow{R_1 \to \infty} \frac{R_1}{\sqrt{\tau_{2}}}\left(\eta\bar{\eta} \right)^{-1} Z_{\Gamma^{17,1}}\xrightarrow{R_2 \to \infty} \frac{R_1 R_2}{\tau_{2}}\left(\eta\bar{\eta} \right)^{-2} Z_{\Gamma^{16}},\\
		&Z_{\Gamma_{\pm}^{18,2}}\xrightarrow{R_2 \to \infty} \frac{R_2}{\sqrt{\tau_{2}}}\left(\eta\bar{\eta} \right)^{-1} Z_{\Gamma_{\pm}^{17,1}|_{(2)}}\xrightarrow{R_1 \to \infty} \frac{R_1 R_2}{\tau_{2}}\left(\eta\bar{\eta} \right)^{-2} Z_{\Gamma^{16}},\\
		&Z_{\Gamma_{\pm}^{18,2}}\xrightarrow{R_1 \to 0}\frac{1}{R_1\sqrt{\tau_{2}}}\left(\eta\bar{\eta} \right)^{-1}Z_{\Gamma_{\pm}^{17,1}|_{(3)}}\xrightarrow{R_2 \to 0} \frac{1}{R_1 R_2\tau_{2}}\left(\eta\bar{\eta} \right)^{-2} Z_{\Gamma^{16}},\\
		&Z_{\Gamma_{\pm}^{18,2}}\xrightarrow{R_2 \to 0}\frac{1}{R_2\sqrt{\tau_{2}}}\left(\eta\bar{\eta} \right)^{-1}Z_{\Gamma^{17,1}}\xrightarrow{R_1 \to 0} \frac{1}{R_1 R_2\tau_{2}}\left(\eta\bar{\eta} \right)^{-2} Z_{\Gamma^{16}},\\
		&Z_{\Gamma_{\pm}^{18,2}}\xrightarrow{R_1 \to \infty,R_2 \to 0} \frac{R_1}{R_2 \tau_{2}}\left(\eta\bar{\eta} \right)^{-2} Z_{\Gamma^{16}},\\
		&Z_{\Gamma_{\pm}^{18,2}}\xrightarrow{R_1 \to 0,R_2 \to \infty} \frac{R_2}{R_1 \tau_{2}}\left(\eta\bar{\eta} \right)^{-2} Z_{\Gamma_{\pm}^{16}},
	\end{align*}
	and $Z_{\Gamma_{\pm}^{18,2}+\delta}$ as
	\begin{align*}
		&Z_{\Gamma_{\pm}^{18,2}+\delta}\xrightarrow{R_1 \to \infty}0,\\
		&Z_{\Gamma_{\pm}^{18,2}+\delta}\xrightarrow{R_2 \to \infty} \frac{R_2}{\sqrt{\tau_{2}}}\left(\eta\bar{\eta} \right)^{-1} Z_{\Gamma_{\pm}^{17,1}+\delta|_{(2)}}\xrightarrow{R_1 \to \infty}0,\\
		&Z_{\Gamma_{\pm}^{18,2}+\delta}\xrightarrow{R_1 \to 0}\frac{1}{R_1\sqrt{\tau_{2}}}\left(\eta\bar{\eta} \right)^{-1}Z_{\Gamma_{\pm}^{17,1}+\delta|_{(3)}}\xrightarrow{R_2 \to 0} 0,\\
		&Z_{\Gamma_{\pm}^{18,2}+\delta}\xrightarrow{R_2\to 0} 0,\\
		&Z_{\Gamma_{\pm}^{18,2}+\delta}\xrightarrow{R_1 \to 0,R_2 \to \infty} \frac{R_2}{R_1 \tau_{2}}\left(\eta\bar{\eta} \right)^{-2} Z_{\Gamma_{\pm}^{16}+\frac{\hat{\pi}}{2}}.
	\end{align*}
	Then the limit $R_{1} \to 0$ gives the 9-dimensional interpolating model, which belongs to the class (3): $R_{2} \to \infty$ gives a 10-dimensional non-supersymmetric model while $R_{2} \to 0$ gives a 10-dimensional supersymmetric model. We also find that $R_{2} \to \infty$ limit gives the 9-dimensional interpolating model, which belongs to the class (2): $R_{1} \to \infty$ gives a 10-dimensional supersymmetric model while $R_{1} \to 0$ gives a 10-dimensional non-supersymmetric model.
	
	\item $\left|\hat{\pi} \right|^2=2~~(\text{mod 4}), ~~\left(\hat{m};\hat{n} \right) =(1,0;1,0)$:\\
	In this class, the inner product (\ref{pdelta}) is written as
	\begin{align}
		\delta\cdot p=\frac{1}{2}\left( \hat{\pi}\cdot\pi+m^{1}+n_{1} \right).
	\end{align}
	We can write $\Gamma_{\pm}^{18,2}$ and $\Gamma_{\pm}^{18,2}+\delta$ as 
	\begin{align}
		&\Gamma_{\pm}^{18,2}=\left\lbrace p=Z\tilde{\mathcal{E}}\left| \left(\pi,m,n \right)\in\left( \Gamma^{16}_{\pm},2\mathbb{Z}\times \mathbb{Z},2\mathbb{Z} \times \mathbb{Z}\right) \right.  \right\rbrace
		\nonumber\\&~~~~~~~~~~~~
		\oplus \left\lbrace p=Z\tilde{\mathcal{E}}\left| \left(\pi,m,n \right)\in\left( \Gamma^{16}_{\pm},(2\mathbb{Z}+1)\times\mathbb{Z},(2\mathbb{Z}+1)\times \mathbb{Z}\right)  \right.  \right\rbrace\nonumber\\&~~~~~~~~~~~~
		\oplus \left\lbrace p=Z\tilde{\mathcal{E}}\left| \left(\pi,m,n \right)\in\left( \Gamma^{16}_{\mp},2\mathbb{Z}\times\mathbb{Z},(2\mathbb{Z}+1)\times \mathbb{Z}\right)  \right.  \right\rbrace\nonumber\\&~~~~~~~~~~~~
		\oplus \left\lbrace p=Z\tilde{\mathcal{E}}\left| \left(\pi,m,n \right)\in\left( \Gamma^{16}_{\mp},(2\mathbb{Z}+1)\times\mathbb{Z},2\mathbb{Z}\times \mathbb{Z}\right)  \right.  \right\rbrace,\\
		&\Gamma_{\pm}^{18,2}+\delta=\left\lbrace p=Z\tilde{\mathcal{E}}\left| \left(\pi,m,n \right)\in\left( \Gamma^{16}_{\pm}+\frac{\hat{\pi}}{2},(2\mathbb{Z}+\frac{1}{2})\times \mathbb{Z},(2\mathbb{Z}+\frac{1}{2}) \times \mathbb{Z}\right)  \right.  \right\rbrace
		\nonumber\\&~~~~~~~~~~~~~~~
		\oplus \left\lbrace p=Z\tilde{\mathcal{E}}\left| \left(\pi,m,n \right)\in\left( \Gamma^{16}_{\mp}+\frac{\hat{\pi}}{2},(2\mathbb{Z}-\frac{1}{2})\times \mathbb{Z},(2\mathbb{Z}-\frac{1}{2})\times \mathbb{Z}\right)  \right.  \right\rbrace\nonumber\\&~~~~~~~~~~~~~~~
		\oplus \left\lbrace p=Z\tilde{\mathcal{E}}\left| \left(\pi,m,n \right)\in\left( \Gamma^{16}_{\mp}+\frac{\hat{\pi}}{2},(2\mathbb{Z}+\frac{1}{2})\times \mathbb{Z},(2\mathbb{Z}-\frac{1}{2})\times \mathbb{Z}\right)  \right.  \right\rbrace\nonumber\\&~~~~~~~~~~~~~~~
		\oplus \left\lbrace p=Z\tilde{\mathcal{E}}\left| \left(\pi,m,n \right)\in\left( \Gamma^{16}_{\mp}+\frac{\hat{\pi}}{2},(2\mathbb{Z}-\frac{1}{2})\times \mathbb{Z},(2\mathbb{Z}+\frac{1}{2})\times \mathbb{Z}\right)  \right.  \right\rbrace.
	\end{align}
	Therefore the behavior of $Z_{\Gamma_{\pm}^{18,2}}$ is
	\begin{align*}
		&Z_{\Gamma_{\pm}^{18,2}}\xrightarrow{R_1 \to \infty} \frac{R_1}{\sqrt{\tau_{2}}}\left(\eta\bar{\eta} \right)^{-1} Z_{\Gamma^{17,1}}\xrightarrow{R_2 \to \infty} \frac{R_1 R_2}{\tau_{2}}\left(\eta\bar{\eta} \right)^{-2} Z_{\Gamma^{16}},\\
		&Z_{\Gamma_{\pm}^{18,2}}\xrightarrow{R_2 \to \infty} \frac{R_2}{\sqrt{\tau_{2}}}\left(\eta\bar{\eta} \right)^{-1} Z_{\Gamma_{\pm}^{17,1}|_{(4)}}\xrightarrow{R_1 \to \infty} \frac{R_1 R_2}{\tau_{2}}\left(\eta\bar{\eta} \right)^{-2} Z_{\Gamma^{16}},\\
		&Z_{\Gamma_{\pm}^{18,2}}\xrightarrow{R_1 \to 0}\frac{1}{R_1\sqrt{\tau_{2}}}\left(\eta\bar{\eta} \right)^{-1}Z_{\Gamma^{17,1}}\xrightarrow{R_2 \to 0} \frac{1}{R_1 R_2\tau_{2}}\left(\eta\bar{\eta} \right)^{-2} Z_{\Gamma^{16}},\\
		&Z_{\Gamma_{\pm}^{18,2}}\xrightarrow{R_2 \to 0}\frac{1}{R_2\sqrt{\tau_{2}}}\left(\eta\bar{\eta} \right)^{-1}Z_{\Gamma_{\pm}^{17,1}|_{(4)}}\xrightarrow{R_1 \to 0} \frac{1}{R_1 R_2\tau_{2}}\left(\eta\bar{\eta} \right)^{-2} Z_{\Gamma^{16}},\\
		&Z_{\Gamma_{\pm}^{18,2}}\xrightarrow{R_1 \to \infty,R_2 \to 0} \frac{R_1}{R_2 \tau_{2}}\left(\eta\bar{\eta} \right)^{-2} Z_{\Gamma^{16}},\\
		&Z_{\Gamma_{\pm}^{18,2}}\xrightarrow{R_1 \to 0,R_2 \to \infty} \frac{R_2}{R_1 \tau_{2}}\left(\eta\bar{\eta} \right)^{-2} Z_{\Gamma^{16}},
	\end{align*}
	and that of $Z_{\Gamma_{\pm}^{18,2}+\delta}$ is
	\begin{align*}
		&Z_{\Gamma_{\pm}^{18,2}+\delta}\xrightarrow{R_1 \to \infty}0,\\
		&Z_{\Gamma_{\pm}^{18,2}+\delta}\xrightarrow{R_2 \to \infty} \frac{R_2}{\sqrt{\tau_{2}}}\left(\eta\bar{\eta} \right)^{-1} Z_{\Gamma_{\pm}^{17,1}+\delta|_{(4)}}\xrightarrow{R_1 \to \infty}0,\\
		&Z_{\Gamma_{\pm}^{18,2}+\delta}\xrightarrow{R_1 \to 0} 0,\\
		&Z_{\Gamma_{\pm}^{18,2}+\delta}\xrightarrow{R_2\to 0}\frac{1}{R_2\sqrt{\tau_{2}}}\left(\eta\bar{\eta} \right)^{-1}Z_{\Gamma_{\pm}^{17,1}+\delta|_{(4)}}\xrightarrow{R_1 \to \infty ~\text{or}~ 0} 0.
	\end{align*}
	In this class, the 9-dimensional supersymmetric models are obtained in both $R_{1} \to \infty$ and $R_{1} \to 0$ limits, and both $R_{2} \to \infty$ and $R_{2} \to 0$ limits give the 9-dimensional non-supersymmetric model which belongs to the class (4). Note that we have the same four endpoint models, so there is no interpolation between the supersymmetric model and the non-supersymmetric one in 10 dimensions.
\end{enumerate}

We end this subsection by further explaining the behavior in the endpoint limits. From the above analysis, we find that if $\hat{m}^{i}~(\hat{n}_{i})=1$ for $i=1,2$ in $d=2$ case, the limit of $R_{i}\to\infty~(0)$ gives the 9-dimensinal supersymmetric model. This behavior is similar to the one in the $d=1$ case that the limit of $R_{1}\to\infty~(0)$ gives the 10-dimensinal supersymmetric model with $\hat{m}^{1}~(\hat{n}_{1})=1$. Moreover, whether the 10-dimensional models obtained by the limit of $R_{1},R_{2}\to\infty~\text{or}~0$ are supersymmetric or not depends on the sum of $\hat{m}^{i}, \hat{n}_{i}$. The conditions to obtain the 10-dimensional (non-)supersymmetric models are shown in Table \ref{table2}. These discussions can be easily extended to the general $d$-dimensionally compactified case.

\begin{table}[h]
	\centering
	\begin{tabular}{|c||c|c|c|}  \hline
		Limits of $R_{1},R_{2}$ & 10D SUSY model & 10D Non-SUSY model \\ \hline \hline
		$(R_{1},R_{2})\to(\infty,\infty)$ & $\hat{m}^{1}+\hat{m}^{2}>0$ & $\hat{m}^{1}+\hat{m}^{2}=0$ \\ \hline
		$(R_{1},R_{2})\to(\infty,0)$ & $\hat{m}^{1}+\hat{n}_{2}>0$ & $\hat{m}^{1}+\hat{n}_{2}=0$ \\ \hline
		$(R_{1},R_{2})\to(0,\infty)$ & $\hat{n}_{1}+\hat{m}^{2}>0$ & $\hat{n}_{1}+\hat{m}^{2}=0$ \\ \hline
		$(R_{1},R_{2})\to(0,0)$ & $\hat{n}_{1}+\hat{n}_{2}>0$& $\hat{n}_{1}+\hat{n}_{2}=0$ \\ \hline
	\end{tabular}
	\caption{The conditions on $\hat{m}^{i},\hat{n}_{i}$ $(i=1,2)$ which determine whether the 10-dimensional models obtained by the limit of $R_{1},R_{2}\to\infty~\text{or}~0$ are supersymmetric or not.} \label{table2}
\end{table}
\section{Massless spectrum}\label{massless spectrum}
We study the massless spectra in the non-supersymmetric string model $d$-dimensionally compactified with general $\mathbb{Z}_{2}$ twists. The analysis in this section, generalized from \cite{Itoyama:2021fwc}, is needed to evaluate the cosmological constant in section \ref{CosmConst}. 

The left- and right-moving mass formulae for heterotic models are given by
\begin{subequations}\label{mass formula}
	\begin{align}
		M^2_{L}&=|\ell_{L}|^{2}+|p_{L}|^{2}+2(N_{L}-1),\\
		M^2_{R}&=|p_{R}|^{2}+2(N_{R}-a_{R}),
	\end{align}
\end{subequations}
where $a_{R}=1/2$ for NS-sector and $a_{R}=0$ for R-sector, and the physical states satisfy the level-matching condition $M^2_{L}=M^2_{R}$. We find two possibilities to get the massless states from (\ref{mass formula}). The first possibility arises from the states which satisfy
\begin{align}\label{sec1}
	N_{L}=1, ~~N_{R}=a_{R},~~|\ell_{L}|=|p_{L}|=|p_{R}|=0.
\end{align}
We call a set of the massless states satisfying this condition sector 1. By using (\ref{internal momenta}), the condition (\ref{sec1}) leads to
\begin{align}
	\pi=m=n=0.
\end{align} 
This condition does not rely on the moduli, so there are always the massless states with (\ref{sec1}) at any point in the moduli space. These massless states consist of a gravity multiplet (a graviton, an anti-symmetric two-form, and dilaton) and gauge bosons of $U(1)^{16}_{L}\times U(1)_{l}\times U(1)_{r}$, whose degrees of freedom are $8\times8$ and $8\times16$ respectively.

The other possibility comes from the states which satisfy
\begin{align}\label{sec2}
	N_{L}=0, ~~N_{R}=a_{R},~~|\ell_{L}|^{2}+|p_{L}|^{2}=2,~~|p_{R}|=0.
\end{align}
We call a set of the massless states satisfying this condition sector 2. Toward evaluating the cosmological constant, we only focus on the zero-winding states with $m=0$. Then, from (\ref{internal momenta}) the condition can be written as
\begin{align}\label{massless cond}
	n_{i}=-\pi\cdot A_{i},~~|\pi|^{2}=2,
\end{align}
where the first condition holds for all $i=1,\ldots,d$, and the second one implies that the massless states correspond to nonzero roots of semisimple subgroup $g'\subset g$ with $g$ being $SO(32)$ or $E_{8}\times E_{8}$. Now we define $\Delta_{g'}$ as a set of nonzero roots of semisimple subgroup $g'\subset g$. For $g'=g$ case,
\begin{align}\label{Delta_SO(32)}
	\Delta_{SO(32)}&=\left\lbrace\left(\underline{\pm,\pm,0^{14}}\right)\right\rbrace,\\
	\Delta_{E_{8}\times E_{8}}&=\left\lbrace\left(\underline{\pm,\pm,0^{6}};0^{8}\right), \frac{1}{2}\left(\underline{\pm,\pm,\pm,\pm,\pm,\pm,\pm,\pm}_{+};0^{8}\right)\right\rbrace\nonumber\\ \label{Delta_E_8E_8}
	&~~~~~~~~~+\left\lbrace\left(0^{8};\underline{\pm,\pm,0^{6}}\right), \frac{1}{2}\left(0^{8};\underline{\pm,\pm,\pm,\pm,\pm,\pm,\pm,\pm}_{+}\right)\right\rbrace,
\end{align}
where the underline indicates permutations of the components, and the subscript $+$ $(-)$ denotes the number of $+$ is even (odd).

As in \cite{Itoyama:2021fwc}, we shall clarify the difference between the toroidal models and the non-supersymmetric models. In the toroidal models, $m\in\mathbb{Z}^{d}, n\in\mathbb{Z}^{d}$ and $\pi\in\Gamma^{16}$ for both spacetime vectors and spinors. From the condition (\ref{massless cond}), we find that in toroidal models, a non-Abelian part of the gauge group is $g'$ if the Wilson lines $A_{i(T)}=A_{i(T)}^{(g')}$ satisfy
\begin{equation}\label{massless_cond_toroidal}
	\begin{cases}
		\pi\cdot A_{i(T)}^{(g')}\in\mathbb{Z} & \text{for}~~\pi\in\Delta_{g'}, \\
		\pi\cdot A_{i(T)}^{(g')}\notin\mathbb{Z} & \text{for}~~\pi\in\Delta_{g}\backslash\Delta_{g'},
	\end{cases}
\end{equation}
where the subscript $T$ means the toroidal models.

In the non-supersymmetric cases, we focus on simplified $d$-dimensinally compactified models. We split the $d$ compactified dimensions into 4 groups: $i=a+b=a_{(2)}+a_{(4)}+b_{(3)}+b_{(1)}$ for $i=1,\ldots,d$, where $a, b$ and $a_{(2)}, a_{(4)}, b_{(3)}, b_{(1)}$ is defined to run over as follows:
\begin{align*}
	a&=1,\ldots,D; ~~~~~~~a_{(2)}=1,\ldots,D_{2}, ~~~~~~~~~~~~~~a_{(4)}=D_{2}+1,\ldots,D,\\
	b&=D+1,\ldots,d; ~~b_{(3)}=D+1,\ldots,D+D_{3}, ~~b_{(1)}=D+D_{3}+1,\ldots,d,
\end{align*}
and we impose the following conditions on $(\hat{m};\hat{n})\in\mathbb{Z}^{2d}$:
\begin{subequations}\label{condition for mn}
	\begin{align}
		&(\hat{m}^{a_{(2)}},\hat{n}_{a_{(2)}})=(1,0),\\
		&(\hat{m}^{a_{(4)}},\hat{n}_{a_{(4)}})=(1,1),\\
		&(\hat{m}^{b_{(3)}},\hat{n}_{b_{(3)}})=(0,1),\\
		&(\hat{m}^{b_{(1)}},\hat{n}_{b_{(1)}})=(0,0).
	\end{align}
\end{subequations}
Here the subscptions $(1),\ldots,(4)$ correspond to the four 9D classes. From (\ref{condition for mn}), we can write down $(\hat{m};\hat{n})$ explicitly as
\begin{align}\label{mhatnhat}
	(\hat{m};\hat{n})=(1^{D}, 0^{d-D}; 0^{D_{2}}, 1^{D+D_{3}-D_{2}}, 0^{d-D-D_{3}}).
\end{align}
The inner product  $\delta\cdot p$ is given by 
\begin{align}
	\delta\cdot p=\frac{1}{2}\left( \pi\cdot\hat{\pi}+\hat{m}n^{t}+\hat{n}m^{t} \right)=\frac{1}{2}\left( \pi\cdot\hat{\pi}+\sum_{a=1}^{D}n_{a}+\sum_{i=D_{2}+1}^{D+D_{3}}m^{i} \right).
\end{align}
Then $\Gamma_{\pm}^{16+d,d}$ is written as the following sets:
\begin{align}\label{Gamma_in_d}
	&\Gamma_{\pm}^{16+d,d}=\left\lbrace p=Z\tilde{\mathcal{E}}\left|  \left(\pi,\sum_{i=D_{2}+1}^{D+D_{3}}m^{i},\sum_{a=1}^{D}n_{a} \right)\in \left(\Gamma^{16}_{\pm},2\mathbb{Z},2\mathbb{Z} \right)  \right.  \right\rbrace
	\nonumber\\&~~~~~~~~~~~~~~~
	\oplus\left\lbrace p=Z\tilde{\mathcal{E}} \left|  \left(\pi,\sum_{i=D_{2}+1}^{D+D_{3}}m^{i},\sum_{a=1}^{D}n_{a} \right)\in\left( \Gamma^{16}_{\pm},2\mathbb{Z}+1,2\mathbb{Z}+1\right) \right.  \right\rbrace
	\nonumber\\&~~~~~~~~~~~~~~~
	\oplus\left\lbrace p=Z\tilde{\mathcal{E}} \left|  \left(\pi,\sum_{i=D_{2}+1}^{D+D_{3}}m^{i},\sum_{a=1}^{D}n_{a} \right)\in\left( \Gamma^{16}_{\mp},2\mathbb{Z},2\mathbb{Z}+1\right) \right.  \right\rbrace
	\nonumber\\&~~~~~~~~~~~~~~~
	\oplus\left\lbrace p=Z\tilde{\mathcal{E}} \left|  \left(\pi,\sum_{i=D_{2}+1}^{D+D_{3}}m^{i},\sum_{a=1}^{D}n_{a} \right)\in\left( \Gamma^{16}_{\mp},2\mathbb{Z}+1,2\mathbb{Z}\right) \right.  \right\rbrace.
\end{align}
Note that $n_{a}$ is either even or odd, which implies that (\ref{massless cond}) leads to $\pi\cdot A_{a}\in2\mathbb{Z}$ or $\pi\cdot A_{a}\in2\mathbb{Z}+1$. 

Let $\Delta_{g}^{\pm}$ denote subsets of $\pi\in\Gamma_{\pm}^{16}$ with $|\pi|^{2}=2$. From (\ref{non-susy hetero}), we notice that the massless vectors and spinors live in $\Gamma_{+}^{16+d,d}$ and $\Gamma_{-}^{16+d,d}$ respectively. Therefore the condition (\ref{massless cond}) implies that for massless vectors,
\begin{align}\label{massless_cond_vec}
	\sum_{a=1}^{D}\pi\cdot A_{a}\in2\mathbb{Z},~\pi\cdot A_{b}\in\mathbb{Z}~\text{for}~\pi\in\Delta_{g}^{+}~\text{and/or}~\sum_{a=1}^{D}\pi\cdot A_{a}\in2\mathbb{Z}+1,~\pi\cdot A_{b}\in\mathbb{Z}~~\text{for}~\pi\in\Delta_{g}^{-},
\end{align}
while for massless spinors,
\begin{align}\label{massless_cond_spi}
	\sum_{a=1}^{D}\pi\cdot A_{a}\in2\mathbb{Z},~\pi\cdot A_{b}\in\mathbb{Z}~\text{for}~\pi\in\Delta_{g}^{-}~\text{and/or}~\sum_{a=1}^{D}\pi\cdot A_{a}\in2\mathbb{Z}+1,~\pi\cdot A_{b}\in\mathbb{Z}~~\text{for}~\pi\in\Delta_{g}^{+},
\end{align}
From the massless conditions (\ref{massless_cond_toroidal}) and (\ref{massless_cond_vec}), we find that $A_{i}^{(g')}$ which realize a gauge group $g'$ in the non-supersymmetric models are expressed in terms of $A_{i(T)}^{(g')}$ in the toroidal models as follows:
\begin{align}\label{WLrelation}
	\sum_{a=1}^{D}A_{a}^{(g')}=2\sum_{a=1}^{D}A_{a(T)}^{(g')}+\hat{\pi}, ~~A_{b}^{(g')}=A_{b(T)}^{(g')}.
\end{align}
This relation makes the conditions (\ref{massless_cond_vec}) for the massless vectors expressed in terms of $A_{i(T)}$ as
\begin{align}\label{massless_cond_vec_toroidal}
	\sum_{a=1}^{D}\pi\cdot A_{a(T)}\in\mathbb{Z},~\pi\cdot A_{b(T)}\in\mathbb{Z}~~\text{for}~\pi\in\Delta_{g'},
\end{align}
while the condition (\ref{massless_cond_spi}) for the massless spinors as
\begin{align}\label{massless_cond_spi_toroidal}
	\sum_{a=1}^{D}\pi\cdot A_{a(T)}\in\mathbb{Z}+\frac{1}{2},~\pi\cdot A_{b(T)}\in\mathbb{Z}~~\text{for}~\pi\in\Delta_{g'}.
\end{align}
These two conditions do not depend on $\hat{\pi}$. Therefore by using $A_{i(T)}$ but not $A_{i}$, we can identify the massless spectrum with $m=0$ in the non-supersymmetric models without specifying the choice of $\hat{\pi}$.

We can specify the massless spectra with $m=0$ by using the conditions discussed above. However, in order to evaluate the cosmological constant, we further discuss the massless conditions (\ref{massless_cond_vec}), (\ref{massless_cond_spi}) and (\ref{massless_cond_vec_toroidal}), (\ref{massless_cond_spi_toroidal}), and the Wilson-line relation (\ref{WLrelation}).

Recall that the massless condition (\ref{massless cond}) gives $\pi\cdot A_{a}\in2\mathbb{Z}$ or $\pi\cdot A_{a}\in2\mathbb{Z}+1$, then the condition for massless vectors (\ref{massless_cond_vec}), in addtion to $\pi\cdot A_{b}\in\mathbb{Z}$, can be written as 
\begin{align}\label{massless_cond_vec2}
	\sum_{a=1}^{D}(2n_{a}-1)(\pi \cdot A_{a})\in2\mathbb{Z}~\text{for}~\pi\in\Delta_{g}^{+}~\text{and/or}~\sum_{a=1}^{D}(2n_{a}-1)(\pi \cdot A_{a})\in2\mathbb{Z}+1~\text{for}~\pi\in\Delta_{g}^{-},
\end{align}
while for massless spinors (\ref{massless_cond_spi}) as
\begin{align}\label{massless_cond_spi2}
	\sum_{a=1}^{D}(2n_{a}-1)(\pi \cdot A_{a})\in2\mathbb{Z}~\text{for}~\pi\in\Delta_{g}^{-}~\text{and/or}~\sum_{a=1}^{D}(2n_{a}-1)(\pi \cdot A_{a})\in2\mathbb{Z}+1~\text{for}~\pi\in\Delta_{g}^{+}.
\end{align}
We can also get the relation on the Wilson lines from (\ref{WLrelation}) as follows:
\begin{equation}\label{WLrelation2}
	\sum_{a=1}^{D}(2n_{a}-1)A_{a}=2\sum_{a=1}^{D}(2n_{a}-1)A_{a(T)}+\hat{\pi}, ~~A_{b}=A_{b(T)}.
\end{equation}
From this relation we get the condition expressed with $A_{i(T)}$ for massless vectors as
\begin{align}\label{massless_cond_vec_toroidal2}
	\sum_{a=1}^{D}(2n_{a}-1)(\pi\cdot A_{a(T)})\in\mathbb{Z},~\pi\cdot A_{b(T)}\in\mathbb{Z}~~\text{for}~\pi\in\Delta_{g'},
\end{align}
while for the massless spinors as
\begin{align}\label{massless_cond_spi_toroidal2}
	\sum_{a=1}^{D}(2n_{a}-1)(\pi\cdot A_{a(T)})\in\mathbb{Z}+\frac{1}{2},~\pi\cdot A_{b(T)}\in\mathbb{Z}~~\text{for}~\pi\in\Delta_{g'}.
\end{align}

\section{Cosmological constant}\label{CosmConst}
We evaluate the one-loop cosmological constant of the $d$-dimensionally compactified non-supersymmetric heterotic models with general $\mathbb{Z}_{2}$ twists in the region where supersymmetry is asymptotically restored in the way used in \cite{Itoyama:1986ei,Itoyama:1987rc,Itoyama:2019yst,Itoyama:2020ifw,Itoyama:2021fwc,Abel:2015oxa,Abel:2017vos,Abel:2020ldo,Angelantonj:2019gvi,Kounnas:2015yrc,Kounnas:2017mad,Faraggi:2009xy,Faraggi:2021mws,Florakis:2016ani,Coudarchet:2017pie,Coudarchet:2018ztz,Coudarchet:2020sjw,Coudarchet:2021qwc}. We show the result does not rely on all the other endpoints. We find the points in the Wilson-line moduli space where the cosmological constant is exponentially suppressed\cite{Kounnas:2016gmz,Kounnas:2017mad, Coudarchet:2018ztz}. We also analyze the Wilson-line moduli stability of the cosmological constant.
\subsection{The formula for the cosmological constant}
The one-loop cosmological constant is given by 
\begin{equation}
	\Lambda^{(10-d)}=-\frac{1}{2}(2\pi \sqrt{\alpha'})^{-(10-d)}\int_{\mathcal{F}}\frac{d^{2}\tau}{\tau_{2}^{2}}Z^{\cancel{SUSY}}_{(\hat{Z})},
\end{equation}
where $\mathcal{F}$ is a fundamental region of the modular group:
\begin{align}
	\mathcal{F}=\left\lbrace \tau=\tau_{1}+i\tau_{2}\in\mathbb{C} \left|~ -\frac{1}{2}\leq\tau_{1}\leq\frac{1}{2},~|\tau|\geq 1 \right\rbrace \right. .
\end{align}

For simplicity, we evaluate the one-loop cosmological constant of the models with $D\geq1$ constructed in section \ref{massless spectrum}, in the region that all $R_{i}  \approx \infty~(i=1,2,\ldots, d)$ where supersymmetry is asymptotically restored. The four classes shown in subsection \ref{d=2 case} are examples we now consider in 8-dimensional models (the $R_1, R_2\to\infty$ limit gives a 10-dimensional supersymmetric model). In that region, only zero-winding states with $m^{i}=0$ contribute to the cosmological constant, and the contribution from the twisted sector is exponentially suppressed. Using Jacobi's abstruse identity $V_8-S_8=0$, the partition function of the Non-supersymmetric heterotic model is written as
\begin{equation}
	Z^{\cancel{SUSY}}_{(\hat{Z})}(\bm{R};\bm{A})\sim Z_{B}^{(8-d)}  \bar{V}_{8} 
	\left\lbrace Z_{\Gamma^{16+d,d}_{+}}(\bm{R};\bm{A})- Z_{\Gamma^{16+d,d}_{-}}(\bm{R};\bm{A})\right\rbrace,
\end{equation}
where $(\bm{R};\bm{A})$ denotes two types of the moduli, radii of the compactified circles and Wilson lines: $(R_{1},\ldots, R_{d};A_{1},\ldots,A_{d})$. The internal part $I=Z_{\Gamma^{16+d,d}_{+}}(\bm{R};\bm{A})- Z_{\Gamma^{16+d,d}_{-}}(\bm{R};\bm{A})$ is written as\footnote{Please do not confuse the lattice element with the ratio of the circumference of a circle to its diameter.}
\begin{equation*}
	I \sim \eta^{-16}(\eta \bar{\eta})^{-d} \sum_{\epsilon=\pm}\epsilon \sum_{\pi \in \Gamma_{\epsilon}^{16}} q^{\frac{|\pi|^2}{2}}
	\prod_{a=1}^{D} \left( \sum_{n_{a} \in 2\mathbb{Z}} - \sum_{n_{a} \in 2\mathbb{Z}+1}\right) e^{-\frac{\pi\tau_{2}}{R_{a}^{2}}(n_{a}+\pi\cdot A_{a})^2}
	\prod_{b=D+1}^{d} \sum_{n_{b} \in \mathbb{Z}} e^{-\frac{\pi\tau_{2}}{R_{b}^{2}}(n_{b}+\pi\cdot A_{b})^2}.
\end{equation*}
Using Poisson's resummation formula,
\begin{align*}
	\left( \sum_{n_{a} \in 2\mathbb{Z}} - \sum_{n_{a} \in 2\mathbb{Z}+1}\right) e^{-\frac{\pi\tau_{2}}{R_{a}^{2}}(n_{a}+\pi\cdot A_{a})^2}
	&=\frac{R_{a}}{\sqrt{\tau_{2}}}\sum_{n_{a} \in \mathbb{Z}}  e^{-\frac{\pi R_{a}^{2}}{4\tau_{2}}(2n_{a}-1)^{2}}e^{\pi i (2n_{a}-1) (\pi \cdot A_{a})}.\\
	\sum_{n_{b} \in \mathbb{Z}} e^{-\frac{\pi\tau_{2}}{R_{b}^{2}}(n_{b}+\pi\cdot A_{b})^2} &= \frac{R_{b}}{\sqrt{\tau_{2}}}\sum_{n_{b}\in \mathbb{Z}}  e^{-\frac{\pi R_{b}^{2}}{\tau_{2}}n_{b}^{2}} e^{2\pi i n_{b} (\pi \cdot A_{b})}.
\end{align*}
Then the partition function is written as
\begin{align}
	Z^{\cancel{SUSY}}_{(\hat{Z})}(\bm{R};\bm{A}) &\sim \prod_{i=1}^{d}R_{i}\times \frac{1}{\tau_{2}^{4}}\eta^{-24}\bar{\eta}^{-8}\bar{V}_{8} 
	\sum_{\epsilon=\pm}\epsilon \sum_{\pi \in \Gamma_{\epsilon}^{16}} q^{\frac{|\pi|^2}{2}}\nonumber \\
	&~~~~\times\sum_{\bm{n}}\exp\left[\pi i \left\lbrace \sum_{a=1}^{D}(2n_{a}-1) (\pi \cdot A_{a})+ \sum_{b=D+1}^{d}(2n_{b})(\pi \cdot A_{b})  \right\rbrace\right]\nonumber \\ &~~~~~~~~~~\times \exp\left[ -\frac{\pi}{4\tau_{2}}\left\lbrace \sum_{a=1}^{D} (2n_{a}-1)^{2}R_{a}^{2} + \sum_{b=D+1}^{d} (2n_{b})^{2}R_{b}^{2} \right\rbrace \right],
\end{align}
where we define $\sum\limits_{\bm{n}}=\prod\limits_{i=1}^{d}\sum\limits_{n_{i}=-\infty}^\infty$. Expanding $\eta^{-24}\bar{\eta}^{-8}\bar{V}_{8} 
\sum\limits_{\pi \in \Gamma_{\epsilon}^{16}} q^{\frac{|\pi|^2}{2}}$ by $q,\bar{q}$ and performing the integral over $\mathcal{F}$ as discussed in \cite{Itoyama:1986ei, Itoyama:1987rc, Abel:2015oxa, Itoyama:2020ifw}, the cosmological constant can be expressed up to exponential suppressed terms as follows:
\begin{align}\label{CC}
	\Lambda^{(10-d)} &\sim -\frac{4! \cdot 2^{d-1}}{\pi^{15-d}(\sqrt{\alpha'})^{10-d}}\left(\prod_{i=1}^{d}R_{i}\right)\sum_{\bm{n}}\left\lbrace \sum_{a=1}^{D} (2n_{a}-1)^{2}R_{a}^{2} + \sum_{b=D+1}^{d} (2n_{b})^{2}R_{b}^{2} \right\rbrace^{-5}\nonumber \\
	&~~~~~~~\times 8\left( 24+\sum_{\epsilon=\pm}\sum_{\pi\in\Delta_{g}^{\epsilon}}\epsilon\exp\left[\pi i \left\lbrace \sum_{a=1}^{D}(2n_{a}-1) (\pi \cdot A_{a})+ \sum_{b=D+1}^{d}(2n_{b})(\pi \cdot A_{b})  \right\rbrace\right] \right),
\end{align}
where $8\times24$ comes from $\pi=0$, which corresponds to the massless states in sector 1. 
To show the cosmological constant (\ref{CC}) is proportional to $n_{F}-n_{B}$, where $n_{B}$ and $n_{F}$ are the degrees of freedom of massless bosons and fermions respectively, we assume that $A_{i}$ satisfy $\pi\cdot A_{i}\in\mathbb{Z}$ for all $\pi$. We can notice that
\begin{equation*}
	\epsilon\exp\left[\pi i \left\lbrace \sum_{a=1}^{D}(2n_{a}-1) (\pi \cdot A_{a})+ \sum_{b=D+1}^{d}(2n_{b})(\pi \cdot A_{b})  \right\rbrace\right] = 
	\begin{cases}
		+1  &   \text{for $A_{i}$ with (\ref{massless_cond_vec2})} \\
		-1  &   \text{for $A_{i}$ with (\ref{massless_cond_spi2})}
	\end{cases}.
\end{equation*}
Recall that the conditions for massless vectors and massless spinors in sector 2 are given by (\ref{massless_cond_vec2}) and (\ref{massless_cond_spi2}) respectively, so this factor assigns $+1$ to massless vectors and $-1$ to massless spinors. Therefore, up to exponential suppressed terms, we get the formula for the cosmological constant, including the contribution $8\times24$ from sector 1 as follows:
\begin{align}\label{formula for cc}
	\Lambda^{(10-d)} \sim \frac{4! \cdot 2^{d-1}}{\pi^{15-d}(\sqrt{\alpha'})^{10-d}}\left( n_{F}-n_{B} \right)\left(\prod_{i=1}^{d}R_{i}\right)\sum_{\bm{n} }\left\lbrace \sum_{a=1}^{D} (2n_{a}-1)^{2}R_{a}^{2} + \sum_{b=D+1}^{d} (2n_{b})^{2}R_{b}^{2} \right\rbrace^{-5}\hspace{-5pt}.
\end{align}
Note that $R_{i}$ are dimensionless radii normalized by $\alpha'$, as we mentioned in subsection \ref{Toroidal model}. The formula (\ref{formula for cc}) shows that if $n_{F}=n_{B}$, which implies the degeneracy between bosons and fermions at massless level, the cosmological constant is exponentially suppressed in the region that supersymmetry is restored. This is the formula generalized for $d$-dimensinally compactified models with general $\mathbb{Z}_{2}$ twists.

We should mention that the cosmological constant (\ref{CC}) is independent of $\hat{\pi}$. Note that
\begin{equation*}
	\exp\left[\pi i (\pi \cdot \hat{\pi}) \right] = 
	\begin{cases}
		+1   &   \text{for $\pi \in \Delta_{g}^{+}$} \\
		-1    &   \text{for $\pi \in \Delta_{g}^{-}$}
	\end{cases},
\end{equation*}
then by using the relation (\ref{WLrelation2}), the cosmological constant (\ref{CC}) can be written as
\begin{align}\label{CC2}
	\Lambda^{(10-d)} &\sim -\frac{4! \cdot 2^{d-1}}{\pi^{15-d}(\sqrt{\alpha'})^{10-d}}\left(\prod_{i=1}^{d}R_{i}\right)\sum_{\bm{n}}\left\lbrace \sum_{a=1}^{D} (2n_{a}-1)^{2}R_{a}^{2} + \sum_{b=D+1}^{d} (2n_{b})^{2}R_{b}^{2} \right\rbrace^{-5}\nonumber \\
	&~\times 8\left( 24+\sum_{\pi\in\Delta_{g}}\exp\left[2\pi i \left\lbrace \sum_{a=1}^{D}(2n_{a}-1) (\pi \cdot A_{a(T)})+ \sum_{b=D+1}^{d}n_{b}(\pi \cdot A_{b(T)})  \right\rbrace\right] \right).
\end{align}
As discussed in \cite{Itoyama:2021fwc}, this does not depend on the choice of $\hat{\pi}$, that is, the choice of all the other endpoint models given by the limit of $R_{i}\to \infty$ or $0$ except for all $R_{i}\to \infty$. We use the expression (\ref{CC2}) to analyze the Wilson-line moduli stability in subsection \ref{stability}. 

We can also get the formula (\ref{formula for cc}) from the expression (\ref{CC2}) by using the conditons (\ref{massless_cond_vec_toroidal2}) and (\ref{massless_cond_spi_toroidal2}). We find that (\ref{formula for cc}) is valid for $A_{i(T)}$ which satisfy $\pi\cdot 2A_{a(T)}\in\mathbb{Z}$ and $\pi\cdot A_{b(T)}\in\mathbb{Z}$ for all $\pi$, since $A_{i(T)}$ with $\sum_{a}\pi\cdot A_{a(T)}\in\mathbb{Z}$ and $\sum_{a}\pi\cdot A_{a(T)}\in\mathbb{Z}+1/2$ give $+1$ and $-1$ contributions respectively.

\subsection{Exponential suppression}
We investigate the massless spectrum with $n_{F}=n_{B}$, where the cosmological constant is exponentially suppressed. In order to use the formula (\ref{formula for cc}), we forcus on the Wilson lines $A_{i(T)}$ with $\pi\cdot 2A_{a(T)}\in\mathbb{Z}$ and $\pi\cdot A_{b(T)}\in\mathbb{Z}$ for all $\pi$.

It is convenient to define $\Delta_{g'}^{(B)}$ and $\Delta_{g'}^{(F)}$ from (\ref{massless_cond_vec_toroidal}) and (\ref{massless_cond_spi_toroidal}) as follows:
\begin{subequations}\label{Delta_B,F}
	\begin{align}
		&\Delta_{g'}^{(B)}=\left\lbrace \pi\in\Delta_{g} \left| ~\sum_{a=1}^{D}\pi\cdot A_{a(T)}\in\mathbb{Z} ~\text{and}~ \pi\cdot A_{b(T)}\in\mathbb{Z} \right. \right\rbrace,\\
		&\Delta_{g'}^{(F)}=\left\lbrace \pi\in\Delta_{g} \left| ~\sum_{a=1}^{D}\pi\cdot A_{a(T)}\in\mathbb{Z}+\frac{1}{2} ~\text{and}~ \pi\cdot A_{b(T)}\in\mathbb{Z} \right. \right\rbrace.
	\end{align}
\end{subequations}
We find that $n_{f}=n_{B}$ is realized if the Wilson lines $A_{i(T)}$ give  the massless states which satisfy the following condition:
\begin{align}\label{massless cond in Delta}
	\left|\Delta_{g'}^{(F)}\right| - \left|\Delta_{g'}^{(B)}\right| =24,
\end{align}
where $\left|\Delta\right|$ implies the number of elements in a set $\Delta$, and $24$ is from the d.o.f. of left-moving massless states in sector 1.

\subsubsection{$Spin(32)/\mathbb{Z}_{2}$ supersymmetric endpoint model}
In this model, $\Delta_{g}=\Delta_{SO(32)}$ given by (\ref{Delta_SO(32)}). As simple configurations, we consider the following Wilson lines $A_{i(T)}$:
\begin{align}\label{WL1}
	A_{a(T)}=\left( 0^{p}, \left(\frac{1}{2}\right)^{q}\right)~\left(p+q=16\right),~~A_{b(T)}=\left( 0^{16}\right).
\end{align}
In other words, we consider that the Wilson lines $A_{a(T)}~(a=1,\ldots, D)$ are the same configuration and $A_{b(T)}~(b=D+1,\ldots, d)$ is taken to be 0. Then we can rewrite this configuration as follows:
\begin{align}
	\sum_{a=1}^{D}A_{a(T)}=\left( 0^{p}, \left(\frac{D}{2}\right)^{q}\right)~\left(p+q=16\right),~~A_{b(T)}=\left( 0^{16}\right).
\end{align}
Note that (\ref{WL1}) satisfies the condition $\pi\cdot 2A_{a(T)}\in\mathbb{Z}$ and $\pi\cdot A_{b(T)}\in\mathbb{Z}$ for all $\pi\in\Delta_{SO(32)}$. In the following, we consider the two cases according to whether $D$ is even or odd.
\begin{enumerate}[(I)]
	\item $D\in2\mathbb{Z}$\\
	From (\ref{Delta_SO(32)}), $\Delta_{g'}^{(B)}=\Delta_{SO(32)}$ and $\Delta_{g'}^{(F)}$ is empty. Therefore we get
	\begin{align}
		\left|\Delta_{g'}^{(F)}\right| - \left|\Delta_{g'}^{(B)}\right| =-480,
	\end{align}
	and we find that there is no solution of (\ref{massless cond in Delta}) in $D\in2\mathbb{Z}$ case. 
	\item $D\in2\mathbb{Z}+1$\\
	In this case, we can write (\ref{Delta_B,F}) as follows:
	\begin{subequations}
		\begin{align}
			&\Delta_{g'}^{(B)}=\left\lbrace \left(\underline{\pm,\pm,0^{p-2}}, 0^{q}\right), \left( 0^{p}, \underline{\pm,\pm,0^{q-2}} \right) \right\rbrace,\\
			&\Delta_{g'}^{(F)}=\left\lbrace \left( \underline{\pm,0^{p-1}}, \underline{\pm,0^{q-1}} \right) \right\rbrace.
		\end{align}
	\end{subequations}
	We find that there are massless gauge bosons transforming in the adjoint representation of $SO(2p)\times SO(2q)$ and massless spinor transforming in $\left(\boldsymbol{2p}, \boldsymbol{2q}\right)$ of $SO(2p)\times SO(2q)$. We can then calculate the number of elements in  $\Delta_{g'}^{(B)}$ and $\Delta_{g'}^{(F)}$ as follows:
	\begin{align}
		\left|\Delta_{g'}^{(F)}\right|-\left|\Delta_{g'}^{(B)}\right| =4pq-\left\lbrace 2p(p-1)+2q(q-1) \right\rbrace.
	\end{align}
	Using $p+q=16$, the solutions of (\ref{massless cond in Delta}) are $(p,q)=\left(\underline{7,9}\right)$. Therefore the cosmological constant is exponentially suppressed when the gauge symmetry is $SO(18)\times SO(14)$ under the configurations of Wilson lines (\ref{WL1}) in $D\in2\mathbb{Z}+1$ case.
\end{enumerate}
\subsubsection{$E_{8}\times E_{8}$ supersymmetric endpoint model}
In this model, $\Delta_{g}=\Delta_{E_{8}\times E_{8}}$ given by (\ref{Delta_E_8E_8}). $\Delta_{E_{8}\times E_{8}}$ is decomposed into two copies of $\Delta_{E_{8}}$ as $\Delta_{E_{8}\times E_{8}} = \Delta_{E_{8}}\oplus\Delta_{E_{8}}$, and $\Delta_{E_{8}}$ is further decomposed as
\begin{equation}
	\Delta_{E_{8}}=\Delta_{SO(16)}\oplus\Delta_{\bm{128}_{+}},
\end{equation}
where $\Delta_{SO(16)}$ and $\Delta_{\bm{128}_{+}}$ are defined as
\begin{equation}
	\Delta_{SO(16)}=\left\lbrace(\underline{\pm,\pm,0^{6}})\right\rbrace,~~~~~\Delta_{\bm{128}_{+}}=\left\lbrace\frac{1}{2}(\underline{\pm,\pm,\pm,\pm,\pm,\pm,\pm,\pm}_{+})\right\rbrace.
\end{equation}
We also notice that $\Delta_{g'}^{(B)}$ and $\Delta_{g'}^{(F)}$ are expressed as the direct sums of two sets as follows:
\begin{align}
	\Delta_{g'}^{(B)}=\Delta_{g_{1}'}^{(B)}+\Delta_{g_{2}'}^{(B)},~~~\Delta_{g'}^{(F)}=\Delta_{g_{1}'}^{(F)}+\Delta_{g_{2}'}^{(F)},
\end{align}
where we define $\Delta_{g_{k}'}^{(B)}$ and $\Delta_{g_{k}'}^{(F)}$ for $\pi=(\pi_{1};\pi_{2})$ and the Wilson lines $A_{a(T)}=\left( A_{1}; A_{2}\right)$ and $A_{b(T)}=\left( A'_{1}; A'_{2}\right)$ as follows:
\begin{subequations}\label{Delta_Bi,Fi}
	\begin{align}
		&\Delta_{g_{k}'}^{(B)}=\left\lbrace \pi_{k}\in\Delta_{E_{8}} \left| ~\sum_{a=1}^{D}\pi_{k}\cdot A_{k}\in\mathbb{Z} ~\text{and}~ \pi_{k}\cdot A'_{k}\in\mathbb{Z}\right. \right\rbrace,\\
		&\Delta_{g_{k}'}^{(F)}=\left\lbrace \pi_{k}\in\Delta_{E_{8}} \left| ~\sum_{a=1}^{D}\pi_{k}\cdot A_{k}\in\mathbb{Z}+\frac{1}{2} ~\text{and}~ \pi_{k}\cdot A'_{k}\in\mathbb{Z}\right. \right\rbrace.
	\end{align}
\end{subequations}
Then the condition (\ref{massless cond in Delta}) can be rewritten as
\begin{align}\label{massless cond in Delta2}
	\sum_{k=1,2}\left(~\left|\Delta_{g_{k}'}^{(F)}\right|-\left|\Delta_{g_{k}'}^{(B)}\right|~\right)=24.
\end{align}
Thus it is sufficient to consider only the first eight components of the Wilson lines and identify $\Delta_{g_{1}'}^{(B)}$ and $\Delta_{g_{1}'}^{(F)}$. 

Let us study the very simple configrations of Wilson lines $A_{a(T)}=\left( A_{1}; A_{2}\right)$ and $A_{b(T)}=\left( A'_{1}; A'_{2}\right)$ as 
\begin{align}\label{WL2}
	A_{k}=\left( 0^{p_{k}}, \left(\frac{1}{2}\right)^{q_{k}} \right)~\left(p_{k}+q_{k}=8\right),~~A'_{k}=\left( 0^{8}\right),~~~\text{for $k=1,2$},
\end{align}
where $q_{k}$ is even so that $\pi\cdot 2A_{a(T)}\in\mathbb{Z}$ for all $\pi$. Here $p_{k}$ and $q_{k}$ are independent of $a$, so the Wilson lines $A_{a(T)}~(a=1,\ldots, D)$ are the same configuration. As in the $Spin(32)/\mathbb{Z}_{2}$ supersymmetric endpoint model, we should consider whether $D$ is even or odd. However, with $D$ even, $\Delta_{g_{1}'}^{(B)}=\Delta_{E_{8}}$ and $\Delta_{g_{1}'}^{(F)}$ is empty, so we obtain
\begin{align}\label{-240}
	\left|\Delta_{g_{1}'}^{(F)}\right|-\left|\Delta_{g_{1}'}^{(B)}\right|=-240.
\end{align}
This means that there is no solution of (\ref{massless cond in Delta}) in $D\in2\mathbb{Z}$. We focus on the $D\in2\mathbb{Z}+1$ case in the following.
\begin{enumerate}
	\item $p_{1}=0$ or $p_{1}=8$\\
	In this case, $\Delta_{g_{1}'}^{(B)}=\Delta_{E_{8}}$ and $\Delta_{g_{1}'}^{(F)}$ has no elements, so we get (\ref{-240}). 
	
	\item $p_{1}=2$ or $p_{1}=6$\\
	For $p_{1}=2$, we find that the following $\pi_{1}\in\Delta_{SO(16)}$ are in $\Delta_{g_{1}'}^{(B)}$ and $\Delta_{g_{1}'}^{(F)}$:
	\begin{subequations}
		\begin{align}
			&\left(\underline{\pm,\pm}, 0^{6}\right), \left( 0^{2}, \underline{\pm,\pm,0^{4}} \right) \in \Delta_{g_{1}'}^{(B)},\\
			&\left( \underline{\pm,0}, \underline{\pm,0^{5}} \right)\in \Delta_{g_{1}'}^{(F)},
		\end{align}
	\end{subequations}
	and the following $\pi_{1}\in\Delta_{\bm{128}_{+}}$ are in $\Delta_{g_{1}'}^{(B)}$ and $\Delta_{g_{1}'}^{(F)}$:
	\begin{subequations}
		\begin{align}
			&\frac{1}{2}(\underline{\pm,\pm}_{-},\underline{\pm,\pm,\pm,\pm,\pm,\pm}_{-}) \in \Delta_{g_{1}'}^{(B)},\\
			&\frac{1}{2}(\underline{\pm,\pm}_{+},\underline{\pm,\pm,\pm,\pm,\pm,\pm}_{+})\in \Delta_{g_{1}'}^{(F)}.
		\end{align}
	\end{subequations}
	Thus we find  that $\Delta_{g_{1}'}^{(B)}$ gives nonzero roots of $SU(2)\times E_{7}$ and $\Delta_{g_{1}'}^{(F)}$ gives $\left(\boldsymbol{2}, \boldsymbol{56}\right)$ of $SU(2)\times E_{7}$. The same results can also be obtained in $p_{1}=6$. We then get 
	\begin{align}\label{-16}
		\left|\Delta_{g_{1}'}^{(F)}\right|-\left|\Delta_{g_{1}'}^{(B)}\right|=-16.
	\end{align}
	
	\item $p_{1}=4$\\
	In this case, we find that the following $\pi_{1}\in\Delta_{SO(16)}$ are in $\Delta_{g_{1}'}^{(B)}$ and $\Delta_{g_{1}'}^{(F)}$:
	\begin{subequations}
		\begin{align}
			&\left(\underline{\pm,\pm,0^{2}}, 0^{4}\right), \left( 0^{4}, \underline{\pm,\pm,0^{2}} \right) \in \Delta_{g_{1}'}^{(B)},\\
			&\left( \underline{\pm,0^{3}}, \underline{\pm,0^{3}} \right)\in \Delta_{g_{1}'}^{(F)},
		\end{align}
	\end{subequations}
	and the following $\pi_{1}\in\Delta_{\bm{128}_{+}}$ are in $\Delta_{g_{1}'}^{(B)}$ and $\Delta_{g_{1}'}^{(F)}$:
	\begin{subequations}
		\begin{align}
			&\frac{1}{2}(\underline{\pm,\pm,\pm,\pm}_{+},\underline{\pm,\pm,\pm,\pm}_{+}) \in \Delta_{g_{1}'}^{(B)},\\
			&\frac{1}{2}(\underline{\pm,\pm,\pm,\pm}_{-},\underline{\pm,\pm,\pm,\pm}_{-})\in \Delta_{g_{1}'}^{(F)}.
		\end{align}
	\end{subequations}
	Therefore we find that $\Delta_{g_{1}'}^{(B)}$ gives nonzero roots of $SO(16)$ and $\Delta_{g_{1}'}^{(F)}$ gives $\boldsymbol{128}$ of $SO(16)$. We then get 
	\begin{align}\label{16}
		\left|\Delta_{g_{1}'}^{(F)}\right|-\left|\Delta_{g_{1}'}^{(B)}\right|=16.
	\end{align}
\end{enumerate}
From (\ref{-240}), (\ref{-16}) and (\ref{16}), we conclude that there is no solution of (\ref{massless cond in Delta2}) in $D\in2\mathbb{Z}+1$ case since no combination of $-240$, $-16$ and $16$ gives $24$.

\subsection{Wilson-line moduli stability}\label{stability}
We analyze the stability of the Wilson-line moduli by using the expression of the cosmological constant ($\ref{CC2}$). Here we omit the subscript $T$ of Wilson lines: $A_{i(T)}\rightarrow A_{i}$.

\subsubsection{$Spin(32)/\mathbb{Z}_{2}$ supersymmetric endpoint model}
Inserting $\Delta_{g}=\Delta_{SO(32)}$ into (\ref{CC2}), the Wilson line dependent part is written as
\begin{align}
	\sum_{\pi\in\Delta_{SO(32)}}\hspace{-5pt}\exp\left[2\pi i \left\lbrace \sum_{a=1}^{D}(2n_{a}-1) (\pi \cdot A_{a})+ \sum_{b=D+1}^{d}n_{b}(\pi \cdot A_{b})  \right\rbrace\right] =4\sum_{I>J}\cos\left[2\pi \theta^{I}\right]\cos\left[2\pi \theta^{J}\right],
\end{align}
where $I,J=1,\ldots,16$ and we define $\theta^{I}$ as
\begin{align}
	\theta^{I}=\sum_{a=1}^{D}(2n_{a}-1)A_{a}^{I}+\sum_{b=D+1}^{d}n_{b}A_{b}^{I}.
\end{align}
We can then get the first derivative of the cosmological constant (\ref{CC2}) as follows:
\begin{align}\label{first derivative1}
	\frac{\partial\Lambda^{(10-d)}}{\partial A_{i}^{I}} &\sim 8\pi\sum_{\bm{n}}C_{\bm{n}} f_{i}\sin\left[2\pi\theta^{I}\right]\sum_{J\neq I}\cos\left[2\pi\theta^{J}\right],
\end{align}
where $f_{i}=2n_{i}-1$ for $i=a$ while $f_{i}=n_{i}$ for $i=b$, and $C_{\bm{n}}$ is a positive prefactor defined as
\begin{align}
	C_{\bm{n}}=\frac{4! \cdot 2^{d+2}}{\pi^{15-d}(\sqrt{\alpha'})^{10-d}}\left(\prod_{i=1}^{d}R_{i}\right)\left\lbrace \sum_{a=1}^{D} (2n_{a}-1)^{2}R_{a}^{2} + \sum_{b=D+1}^{d} (2n_{b})^{2}R_{b}^{2} \right\rbrace^{-5}.
\end{align}
Let us consider the simple Wilson lines given in (\ref{WL1}). Inserting (\ref{WL1}) into (\ref{first derivative1}), we find that the Wilson lines (\ref{WL1}) are critical points for both even and odd $D$:
\begin{align}
	\frac{\partial\Lambda^{(10-d)}}{\partial A_{i}^{I}}\sim 0~~~~(I=1,\ldots,16,~i=1,\ldots,d).
\end{align}
Next, we calculate the second derivative of the cosmological constant, which is the Hessian. From (\ref{first derivative1}), we get
\begin{equation}\label{second derivative1}
	\frac{\partial^{2}\Lambda^{(10-d)}}{\partial A_{i}^{I}\partial A_{j}^{J}}\sim
	\begin{cases}
		-16\pi^{2}\sum_{\bm{n}}C_{\bm{n}} f_{i}f_{j}\sin\left[2\pi \theta^{I}\right]\sin\left[2\pi \theta^{J}\right] & (I\neq J), \\
		16\pi^{2}\sum_{\bm{n}}C_{\bm{n}} f_{i}f_{j}\cos\left[2\pi\theta^{I}\right]\sum_{K\neq I}\cos\left[2\pi\theta^{K}\right] & (I=J).
	\end{cases}
\end{equation}
The second derivatives (\ref{second derivative1}) with $I\neq J$ vanish in the configuration of Wilson lines (\ref{WL1}). For the components with $I=J$, we consider two cases where $D$ is even or odd.
\begin{enumerate}[(I)]
	\item $D\in2\mathbb{Z}$\\
	By using (\ref{WL1}), the second derivatives (\ref{second derivative1}) with $I=J$ are written as
	\begin{align}
		\frac{\partial^{2}\Lambda^{(10-d)}}{\partial A_{i}^{I}\partial A_{j}^{I}}\sim240\pi^{2}\sum_{\bm{n}}C_{\bm{n}} f_{i}f_{j}~~~~(I=1,\ldots,16).
	\end{align}
	We can notice that $\sum_{\bm{n}}C_{\bm{n}} f_{i}f_{j}=0$ for $i\neq j$, and $\sum_{\bm{n}}C_{\bm{n}} f_{i}f_{j}>0$ for $i=j$. Thus the Hessian matrix is positive definite in the configuration of the Wilson lines (\ref{WL1}). Note that $\Lambda^{(10-d)}$ takes a global minimum when the Wilson lines are given by (\ref{WL1}), where the gauge group is $SO(32)$ and no massless fermions exist.
	
	\item $D\in2\mathbb{Z}+1$\\
	In serting (\ref{WL1}) into (\ref{second derivative1}), then we get the second derivatives with $I=J$ as follows:
	\begin{equation}
		\frac{\partial^{2}\Lambda^{(10-d)}}{\partial A_{i}^{I}\partial A_{j}^{I}}\sim
		\begin{cases}
			16\pi^{2}\left(2p-17\right)\sum_{\bm{n}}C_{\bm{n}} f_{i}f_{j} & (I=1,\ldots,p), \\
			16\pi^{2}\left(-2p+15\right)\sum_{\bm{n}}C_{\bm{n}} f_{i}f_{j} & (I=p+1,\ldots,16).
		\end{cases}
	\end{equation}
	We find that the Hessian matrix is positive definite with $p=0~\text{or}~16$, while negative definite with $p=8$. In other words, $\Lambda^{(10-d)}$ takes a global minimum when the gauge group is $SO(32)$ while a local maximum when the gauge group is $SO(16)\times SO(16)$. We also find that the $\Lambda^{(10-d)}$ lies in the saddle points in the case of $p=7~\text{or}~9$, which give the exponentially suppressed cosmological constant.
\end{enumerate}

\subsubsection{$E_{8}\times E_{8}$ supersymmetric endpoint model}
We obtain the Wilson line dependent part by inserting $\Delta_{g}=\Delta_{E_{8}\times E_{8}}$ into (\ref{CC2}) as follows:
\begin{align}\label{CC_E_8}
	&\sum_{\pi\in\Delta_{E_{8}\times E_{8}}}\exp\left[2\pi i \left\lbrace \sum_{a=1}^{D}(2n_{a}-1) (\pi \cdot A_{a})+ \sum_{b=D+1}^{d}n_{b}(\pi \cdot A_{b})  \right\rbrace\right] \nonumber\\
	=&\sum_{k=1,2}\left\{4\sum_{I_{k}>J_{k}}\cos\left[2\pi\theta^{I_{k}}\right]\cos\left[2\pi  \theta^{J_{k}}\right] +128\left(\prod_{I_{k}}\cos\left[\pi \theta^{I_{k}}\right]+\prod_{I_{k}}\sin\left[\pi \theta^{I_{k}}\right]\right)\right\},
\end{align}
where $I_{1},J_{1}=1,\ldots,8$ and $I_{2},J_{2}=9,\ldots,16$, respectively. Note that the sum over $k=1,2$ implies that $\Delta_{E_{8}\times E_{8}}$ is decomposed into two copies of $\Delta_{E_{8}}$, so it is sufficient to consider the half part ($k=1$) of $\Lambda^{(10-d)}$. The first derivative of $\Lambda^{(10-d)}$ can be obtained as 
\begin{align}\label{first derivative2}
	\frac{\partial\Lambda^{(10-d)}}{\partial A_{i}^{I_{1}}} &\sim 8\pi\sum_{\bm{n}}C_{\bm{n}} f_{i}\left\{\sin\left[2\pi\theta^{I_{1}}\right]\sum_{J_{1}\neq I_{1}}\cos\left[2\pi\theta^{J_{1}}\right]\right.\nonumber\\
	&~~~~~~~~~~+16\left.\left(\sin\left[\pi \theta^{I_{1}}\right]\prod_{J_{1}\neq I_{1}}\cos\left[\pi\theta^{J_{1}}\right]-\cos\left[\pi\theta^{I_{1}}\right]\prod_{J_{1}\neq I_{1}}\sin\left[\pi\theta^{J_{1}}\right]\right)\right\}.
\end{align}
We can also get the second derivative of $\Lambda^{(10-d)}$. For $I_{1}\neq J_{1}$,
\begin{align}\label{second derivative2-1}
	&\frac{\partial^{2}\Lambda^{(10-d)}}{\partial A_{i}^{I_{1}}\partial A_{j}^{J_{1}}}\sim -16\pi^{2}\sum_{\bm{n}}C_{\bm{n}} f_{i}f_{j}\Biggl\{\sin\left[2\pi\theta^{I_{1}}\right]\sin\left[2\pi\theta^{J_{1}}\right]\Biggr.\nonumber\\
	&+8\left.\left(\sin\left[\pi \theta^{I_{1}}\right]\sin\left[\pi\theta^{J_{1}}\right]\prod_{K_{1}\neq I_{1},J_{1}}\cos\left[\pi\theta^{K_{1}}\right]+\cos\left[\pi \theta^{I_{1}}\right]\cos\left[\pi\theta^{J_{1}}\right]\prod_{K_{1}\neq I_{1},J_{1}}\sin\left[\pi\theta^{K_{1}}\right]\right)\right\},
\end{align}
and for $I_{1}=J_{1}$,
\begin{align}\label{second derivative2-2}
	&\frac{\partial^{2}\Lambda^{(10-d)}}{\partial A_{i}^{I_{1}}\partial A_{j}^{J_{1}}}\sim
	16\pi^{2}\sum_{\bm{n}}C_{\bm{n}} f_{i}f_{j}\Biggl\{\cos\left[2\pi\theta^{I_{1}}\right]\sum_{K_{1}\neq I_{1}}\cos\left[2\pi\theta^{K_{1}}\right]\Biggr.\nonumber\\
	&~~~~~~~~~~~~~~~~~~~~~~~~~~~~~~~~~~~~~~~~~~~~~~~~+8\left.\left(\prod_{I_{1}=1}^{8}\cos\left[\pi\theta^{I_{1}}\right]+\prod_{I_{1}=1}^{8}\sin\left[\pi\theta^{I_{1}}\right]\right)\right\}.
\end{align}
From now on, we focus on the simple configuration of the Wilson lines (\ref{WL2}). 

\begin{enumerate}[(I)]
	\item $D\in2\mathbb{Z}$\\
	In this case, the first derivative (\ref{first derivative2}) and the second derivative with $I_{1}\neq J_{1}$ (\ref{second derivative2-1}) vanish in (\ref{WL2}). The second derivative with $I_{1}=J_{1}$ (\ref{second derivative2-2}) can be expressed as
	\begin{align}
		\frac{\partial^{2}\Lambda^{(10-d)}}{\partial A_{i}^{I_{1}}\partial A_{j}^{I_{1}}}\sim240\pi^{2}\sum_{\bm{n}}C_{\bm{n}} f_{i}f_{j}~~~~(I_{1}=1,\ldots,8).
	\end{align}
	Therefore, the Hessian matrix is positive definite in the configuration of the Wilson lines (\ref{WL2}). We also find that the half part of $\Lambda^{(10-d)}$ takes a global minimum when the Wilson lines are given by (\ref{WL2}), where the gauge group is $E_{8}$, and there are no massless fermions, as in $Spin(32)/\mathbb{Z}_{2}$ supersymmetric endpoint model with $D$ even.
	
	\item $D\in2\mathbb{Z}+1$\\
	By using (\ref{WL2}), the first derivative (\ref{first derivative2}) vanishes since $p_{1}$ is even, so the Wilson lines (\ref{WL2}) are the critical points of $\Lambda^{(10-d)}$. The second derivative with $I_{1}\neq J_{1}$ (\ref{second derivative2-1}) vanishes for $I_{1}=1,\ldots,p_{1}, J_{1}=p_{1}+1,\ldots,8$ and $I_{1}=p_{1}+1,\ldots,8, J_{1}=1,\ldots,p_{1}$. We also find for $I_{1},J_{1}=1,\ldots,p_{1}$,
	\begin{equation}
		\frac{\partial^{2}\Lambda^{(10-d)}}{\partial A_{i}^{I_{1}}\partial A_{j}^{J_{1}}}\sim
		\begin{cases}
			0 & (p_{1}\neq 2), \\
			-128\pi^{2}\sum_{\bm{n}}C_{\bm{n}} f_{i}f_{j} & (p_{1}=2).
		\end{cases}
	\end{equation}
	and for $I_{1},J_{1}=p_{1}+1,\ldots,8$,
	\begin{equation}
		\frac{\partial^{2}\Lambda^{(10-d)}}{\partial A_{i}^{I_{1}}\partial A_{j}^{J_{1}}}\sim
		\begin{cases}
			0 & (p_{1}\neq 6), \\
			-128\pi^{2}\sum_{\bm{n}}C_{\bm{n}} f_{i}f_{j} & (p_{1}=6).
		\end{cases}
	\end{equation}
	The second derivative with $I_{1}=J_{1}$ (\ref{second derivative2-2}) can be written for $I_{1}=1,\ldots,p_{1}$ as
	\begin{equation}
		\frac{\partial^{2}\Lambda^{(10-d)}}{\partial A_{i}^{I_{1}}\partial A_{j}^{I_{1}}}\sim
		\begin{cases}
			16\pi^{2}(2p_{1}-9)\sum_{\bm{n}}C_{\bm{n}} f_{i}f_{j} & (p_{1}\neq 8), \\
			240\pi^{2}\sum_{\bm{n}}C_{\bm{n}} f_{i}f_{j} & (p_{1}=8),
		\end{cases}
	\end{equation}
	and for $I_{1}=p_{1}+1,\ldots,8$ as
	\begin{equation}
		\frac{\partial^{2}\Lambda^{(10-d)}}{\partial A_{i}^{I_{1}}\partial A_{j}^{I_{1}}}\sim
		\begin{cases}
			16\pi^{2}(-2p_{1}+7)\sum_{\bm{n}}C_{\bm{n}} f_{i}f_{j} & (p_{1}\neq 0), \\
			240\pi^{2}\sum_{\bm{n}}C_{\bm{n}} f_{i}f_{j} & (p_{1}=0).
		\end{cases}
	\end{equation}
	In the configuration of Wilson lines (\ref{WL2}), the Hessian matrix is positive definite with $p_{1}=0~\text{or}~8$, while negative definite with $p_{1}=4$. The Wilson lines with $p_{1}=0~\text{or}~8$ correspond to global minima where the gauge group is $E_{8}$, while ones with $p_{1}=4$ to a local maximum where the gauge group is $SO(16)\times SO(16)$. We also find that the Wilson lines (\ref{WL2}) with $p_{1}=2~\text{or}~6$ give the saddle points of $\Lambda^{(10-d)}$.
	
\end{enumerate}

\section{Summary}
In this paper, we have studied $d$-dimensionally compactified non-supersymmetric heterotic string models, including interpolating models, with general $\mathbb{Z}_{2}$ twists (which means the arbitrary number of freely acting $\mathbb{Z}_{2}$ twisted directions). 

We have first investigated the limits of the compactified radii to zero and infinity (the endpoint limits) in the $d=2$ case. Then, we have focused on the four concrete examples and shown the various pattern of interpolation of these non-supersymmetric heterotic models. From these analyses, we have found that the interpolation patterns in $d=2$ are recognized as the combinations of those in $d=1$. We also found the conditions to obtain the 10-dimensional (non-)supersymmetric models in the endpoint limits. These discussions are easily extended to the $d$-dimensinally compactified models.

Second, we have studied the massless spectra in the non-supersymmetric heterotic string models with general $\mathbb{Z}_{2}$ twists. Comparing non-supersymmetric models with toroidal ones, we have written down the relation between the Wilson lines. The massless conditions and the Wilson-line relation have been used in calculating the one-loop cosmological constant of non-supersymmetric models.

Next, we have evaluated the cosmological constant $\Lambda^{(10-d)}$ in the region that the supersymmetry asymptotically restored. $\Lambda^{(10-d)}$ is the function of the two types of moduli: the compactified radii and the Wilson lines. Under the assumption of the Wilson lines, we have derived the formula which has the factor of $n_{F}-n_{B}$, where $n_{F}\left(n_{B}\right)$ is the d.o.f. of massless fermions (bosons). This formula is the generalized version for $d$-dimensionally compactified models from the one for $9$-dimensional models calculated in \cite{Itoyama:1986ei, Itoyama:1987rc}. We emphasize this formula is not only for interpolating models but also for the non-supersymmetric ones which give the same 10-dimensional supersymmetric strings in the endpoint limits, and does not depend on the choice of all the other endpoint models.

By using the formula, we have found out the configurations of the Wilson-line moduli where the cosmological constant is exponentially suppressed in the $Spin(32)/\mathbb{Z}_{2}$ supersymmetric endpoint model. In contrast, we cannot find the configuration of the Wilson lines, which give the exponentially suppressed cosmological constant in the $E_{8}\times E_{8}$ supersymmetric endpoint model. In addition, the stability of the Wilson-line moduli of the cosmological constant has been analyzed. We have found that the Wilson lines, which give the exponential suppression, correspond to the saddle points of $\Lambda^{(10-d)}$, and the global minimum of $\Lambda^{(10-d)}$ is negative in both models. These results imply that the moduli are unstable when $\Lambda^{(10-d)}$ is zero or positive up to the exponentially suppressed terms, and the points of stable moduli correspond to anti-de Sitter (AdS) vacua. This situation is considered undesirable in the context of string phenomenology. However, since one can construct perturbatively stable solutions on even AdS backgrounds in the non-supersymmetric $SO(16)\times SO(16)$ heterotic strings \cite{Baykara:2022cwj}, it is interesting to generalize the constructions of stable AdS vacua in the general $\mathbb{Z}_{2}$ twists.

We mention that the formula with $n_{F}-n_{B}$ factor is only valid for the Wilson lines such that $\pi\cdot A_{i}\in\mathbb{Z}$ for all $\pi\in\Delta_{g}$, which give the massless states. The Wilson lines that give the massive states may contribute to the leading term of the cosmological constant in the case with the general $\mathbb{Z}_{2}$ twists and may realize the stable de-Sitter vacua. In addition, our discussion and calculation are based on the one-loop level, so it is worth calculating the higher loop corrections of the cosmological constant.

\section*{Acknowledgments}
I would like to thank Hiroshi Itoyama and Sota Nakajima for past collaborations.
This work was supported by JST, the establishment of university fellowships towards the creation of science technology innovation, Grant Number JPMJFS 2138.


\appendix

\section{$SO(2n)$ characters}\label{appendixA}
The $SO(2n)$ characters are defined as follows:
\begin{align}
	O_{2n}
	&=\frac{1}{2\eta^{n}}\left( \vartheta^{n}
	\begin{bmatrix} 
		0\\ 
		0\\ 
	\end{bmatrix}(0,\tau)+ \vartheta^{n}
	\begin{bmatrix} 
		0\\ 
		1/2\\ 
	\end{bmatrix}(0,\tau)
	\right),\\
	V_{2n}
	&=\frac{1}{2\eta^{n}}\left( \vartheta^{n}
	\begin{bmatrix} 
		0\\ 
		0\\ 
	\end{bmatrix}(0,\tau)- \vartheta^{n}
	\begin{bmatrix} 
		0\\ 
		1/2\\ 
	\end{bmatrix}(0,\tau)
	\right),\\
	S_{2n}
	&=\frac{1}{2\eta^{n}}\left( \vartheta^{n}
	\begin{bmatrix} 
		1/2\\ 
		0\\ 
	\end{bmatrix}(0,\tau)+ \vartheta^{n}
	\begin{bmatrix} 
		1/2\\ 
		1/2\\ 
	\end{bmatrix}(0,\tau)
	\right),\\
	C_{2n}
	&=\frac{1}{2\eta^{n}}\left( \vartheta^{n}
	\begin{bmatrix} 
		1/2\\ 
		0\\ 
	\end{bmatrix}(0,\tau)-\vartheta^{n}
	\begin{bmatrix} 
		1/2\\ 
		1/2\\ 
	\end{bmatrix}(0,\tau)
	\right),
\end{align}
where the Dedekind eta function and the theta function with characteristics are defined as
\begin{align}
	\eta(\tau)&=q^{1/24}\prod_{n=1}^{\infty}\left( 1-q^{n}\right),\\
	\vartheta
	\begin{bmatrix} 
		\alpha\\ 
		\beta\\ 
	\end{bmatrix}(z,\tau)&=\sum_{n=-\infty}^{\infty}\exp\left( \pi i (n+\alpha)^2 \tau +2\pi i (n+\alpha)(z+\beta) \right). 
\end{align} 
From this definition, we find the transformations of $SO(2n)$ characters under $T: \tau\to\tau+1$ and $S: \tau\to-\frac{1}{\tau}$ as follows:
\begin{align}
	T:\left( O_{2n}, V_{2n}, S_{2n}, C_{2n} \right)\mathcal{T}_{2n},\\
	S:\left( O_{2n}, V_{2n}, S_{2n}, C_{2n} \right)\mathcal{S}_{2n},
\end{align}
where $\mathcal{T}_{2n}$ and $\mathcal{S}_{2n}$ are represented as
\begin{align}
	\mathcal{T}_{2n}=\left(\begin{array}{cccc}
		e^{-\frac{i\pi n}{12}} & 0 & 0 & 0 \\
		0 & -e^{\frac{i\pi n}{12}} & 0 & 0 \\
		0 & 0 & e^{\frac{i\pi n}{6}} &0 \\
		0 & 0 & 0 & e^{\frac{i\pi n}{6}}
	\end{array}
	\right),~~ \mathcal{S}_{2n}=\left(\begin{array}{cccc}
		1 & 1 & 1 & 1 \\
		1 & 1 & -1 & -1 \\
		1 & -1 & i^{n} &-i^{n} \\
		1 & -1 & -i^{n} & i^{n}
	\end{array}
	\right).
\end{align}
These transformations can be used to determine twisted sectors of the partition functions.
\bibliography{reference.bib} 
\bibliographystyle{utphys.bst}
\end{document}